\documentclass[aps,pre,twocolumn, groupedaddress,nofootinbib,floatfix,longbibliography]{revtex4-1}

\usepackage{graphicx}
\usepackage{float}
\usepackage{xcolor}
\usepackage[normalem]{ulem}

\usepackage{amsmath}   
\usepackage{amssymb}   
\usepackage{amsthm}    
\usepackage{hyperref}
\usepackage{enumerate}
\usepackage{comment}

\extrafloats{10}

\definecolor{blue}{rgb}{0.2,0.6,1}
\definecolor{dark_blue}{rgb}{0,0,0.8}
\definecolor{dark_red}{rgb}{0.5,0,0}
\definecolor{dark_green}{rgb}{0,.5,0}
\definecolor{cyan}{rgb}{0,0.4,0.5}
\definecolor{myGreen}{RGB}{19,132,23}

\newcommand{\fig}{Fig.~}

\newcommand{\temp}{\theta}

\newcommand{\mE}{m^{}_\mathrm{E}}
\newcommand{\mEeq}{m^*_\mathrm{E}}

\newcommand{\xE}{x^{}_\mathrm{E}}

\newcommand{\mA}{m_\mathrm{A}}
\newcommand{\mB}{m^{}_\mathrm{B}}

\newcommand{\fA}{f_\mathrm{A}}
\newcommand{\fB}{f_\mathrm{B}}

\begin{document}


\title{Pattern localization to a domain edge}

\author{Manon C. Wigbers}
\thanks{These three authors contributed equally}
\author{Fridtjof Brauns}
\thanks{These three authors contributed equally}
\author{Tobias Hermann}
\thanks{These three authors contributed equally}
\author{Erwin Frey}
\email{frey@lmu.de}

\affiliation{
Arnold Sommerfeld Center for Theoretical Physics (ASC) and Center for NanoScience (CeNS), Department of Physics, Ludwig-Maximilians-Universit\"at M\"unchen, Theresienstra\ss e 37, D--80333 M\"unchen, Germany
}

\date{\today}

\begin{abstract}
The formation of protein patterns inside cells is generically described by reaction--diffusion models. 
The study of such systems goes back to Turing, who showed how patterns can emerge from a homogenous steady state when two reactive components have different diffusivities (e.g.\ membrane-bound and cytosolic states). 
However, in nature, systems typically develop in a heterogeneous environment, where upstream protein patterns affect the formation of protein patterns downstream.
Examples for this are the polarization of Cdc42 adjacent to the previous bud-site in budding yeast, and the formation of an actin-recruiter ring that forms around a PIP3 domain in macropinocytosis. 
This suggests that previously established protein patterns can serve as a template for downstream proteins and that these downstream proteins can `sense' the edge of the template.
A mechanism for how this edge sensing may work remains elusive.

Here we demonstrate and analyze a generic and robust edge-sensing mechanism, based on a two-component mass-conserving reaction-diffusion (McRD) model. 
Our analysis is rooted in a recently developed theoretical framework for McRD systems, termed local equilibria theory. 
We extend this framework to capture the spatially heterogeneous reaction kinetics due to the template. 
This enables us to graphically construct the stationary patterns in the phase space of the reaction kinetics. 
Furthermore, we show that the protein template can trigger a regional mass-redistribution instability near the template edge, leading to the accumulation of protein mass, which eventually results in a stationary peak at the template edge.
We show that simple geometric criteria on the reactive nullcline's shape predict when this edge-sensing mechanism is operational.
Thus, our results provide guidance for future studies of biological systems, and for the design of synthetic pattern forming systems.
\end{abstract}


\maketitle

\section{Introduction}
\label{sec:intro}

\subsection{Background and motivation}
\label{sec:motivation}
Many cellular processes, such as cell division and cell motility, rely crucially on the localization of proteins in space and time. 
Strikingly, these protein localization patterns can emerge from the collective coordination of transport and local molecular interactions of proteins. 
Diffusion in the cytosol is a simple means of protein transport that accounts for many self-organization processes \cite{Halatek.etal2018}.
To analyze how the interplay of diffusive protein transport and protein-protein interactions on a nanometer scale influences the protein patterns on the cellular scale, mass-conserving reaction--diffusion models have proven useful \cite{Howard.etal2001,Huang.etal2003,Mori.etal2008,Goryachev.Pokhilko2008,Ishihara.etal2007,Otsuji.etal2007,Halatek.Frey2012,Klunder.etal2013,Trong.etal2014,Alonso.Bar2014,Wu.etal2016,Goryachev.Leda2017,Murray.Sourjik2017,Denk.etal2018,Cusseddu.etal2018,Chiou.etal2018,Glock.etal2019}.
The study of reaction--diffusion systems in general goes back to Turing \cite{Turing1952}, who showed how patterns can emerge from a homogenous steady state when two reactive components have different diffusivities. 
In cells, differential diffusivities are generic because many proteins have membrane-bound and cytosolic states, where diffusion on the membrane is orders of magnitude slower than in the cytoplasm.
Turing's pioneering work \cite{Turing1952} has led to vast advances in the field on how protein patterns arise from homogeneous (initial) steady states on spatially homogeneous domains.
However, as Turing already pointed out \cite{Turing1952}, ``most of an organism, most of the time, is developing from one pattern into another, rather than from homogeneity into a pattern.''

For example, previously formed protein patterns can control pattern formation of proteins downstream by affecting their local interactions, such that the upstream pattern acts as a \emph{spatial template} for the downstream proteins.
A biological system where such ``templating'' has been suggested is macropinocytosis \cite{Williams.etal2019}. 
Here, a high density domain of PIP3 (a charged phospholipid) and a Ras-GTPase\footnote{GTPases are hydrolase enzymes that can bind and hydrolyze guanosine triphosphate (GTP). Ras is a subfamily of small \mbox{GTPases}.} have been suggested to serve as a template for a ring of actin recruiters (SCAR complex, Arp2/3), that forms around the PIP3 domain edge \cite{Veltman.etal2016}. 
Recruitment of an actomyosin ring, controlled by GTPases, is also key for single-cell wound healing.
Following the rupture of the cell wall, two GTPases-- Abr and Cdc42 --are recruited to the wound edge, where they organize into two concentric rings of high protein concentration \cite{Vaughan.etal2011}. 
Cdc42 in turn recruits actomyosin which contracts to close the wound and repair the underlying cytoskeleton. 
Mutations of Abr, which forms the inner ring, leads to a loss of the outside Cdc42 ring, suggesting hierarchical interaction between Abr and Cdc42 \cite{Vaughan.etal2011}. 
Thus, the inner Abr-ring may be pictured as a template for the outer Cdc42-ring.
Yet another example where protein patterns act as a spatial template, can be found during cell division in budding yeast.  
Here, landmark proteins direct the polarization of the GTPase Cdc42, such that the Cdc42 cluster emerges either adjacent to the previous bud-site, or at the opposite cell pole, depending on the cell-type \cite{Chant.Pringle1995}.
Various mutations or deletions of individual landmark proteins lead to Cdc42 clusters right on top of the previous bud-site or at a random position \cite{Tong.etal2007,Lo.etal2013,Miller.etal2017}.
Hence, the landmark proteins may be pictured as a template that controls Cdc42 pattern formation. 
Common to all the above examples is that the downstream proteins localize at the edge of some template. 
Both the specific (molecular) mechanisms, and the general principles underlying this `edge sensing' remain elusive.

Here, we present a pattern-forming mechanism capable of robust edge sensing and provide criteria for its operation based on simple geometric relations in the phase space of the reaction kinetics. 
To find these criteria, we use a recently developed framework, termed \emph{local equilibria theory} which enables us to gain insight into the dynamics of \emph{mass-conserving reaction--diffusion} (McRD) systems \cite{Halatek.Frey2018,Brauns.etal2018}.
We briefly review the key elements of the local equilibria theory for a paradigmatic model for cell polarization in Section~\ref{sec:recap}.
We then introduce a step-like template that imposes heterogeneity in the reaction kinetics, and generalize the framework to study the dynamics of such systems.
This enables us to explain why and under which conditions a density peak forms at the edge of the template. 
Thus, our results may provide guidance for the design of patterns in synthetic systems and may help to identify molecular mechanisms underlying edge-sensing in biological systems.

\subsection{Local equilibria theory}
\label{sec:recap}
\begin{figure*}
\includegraphics{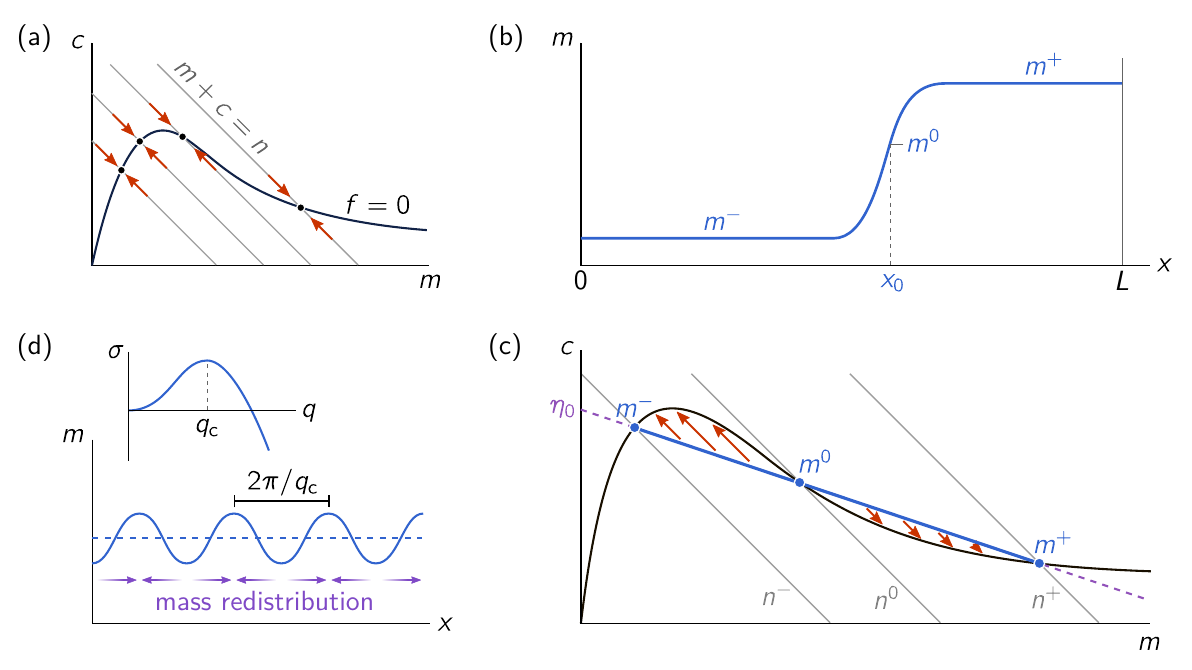}
\caption{\label{fig:no_template_pattern_overview} 
Illustration of the phase-space geometric analysis for two-component McRD systems. 
(a) The reactive equilibria (black dots) are given by the intersections between the reactive subspaces $m+c=n$ (grey lines) and the reactive nullcline $f=0$ (black line) in the $(m,c)$-phase space. 
Hence, the reactive nullcline encodes the qualitative structure of the reactive flow as illustrated by the red arrows.
(b) Sketch of the membrane profile of a mesa pattern composed of a high- and low-density domains, $m^+$ and $m^-$, connected by a diffusive interface around the inflection point $x_0$. 
(c) Flux-balance construction of the mesa pattern in phase space (cf.~(a)). 
The intersections of the flux-balance subspace (FBS) (purple dashed line) and the reactive nullcline yield the concentrations at the plateaus $m^\pm$ and the inflection point $m^0$. 
The balance of net reactive flows in the system (red arrows) determines the FBS-offset $\eta_0$. 
In the regime where the slope of the reactive nullcline is steeper than the slope of the FBS, an homogenous steady state is laterally unstable.
(d) Linearization of the dynamics in the vicinity of the homogeneous steady state, yields a dispersion relation for growth rates of the eigenfunctions (Fourier modes indexed by wavenumber $q$).
The fastest growing mode, $q_c$, dominates the length scale of the initial dynamics.
}
\end{figure*}

We consider the dynamics of one protein species on a one-dimensional domain of length $L$, as in Ref.~\cite{Brauns.etal2018}. 
The proteins can cycle between a membrane-bound state (concentration $m(x,t)$) and a cytosolic state (concentration $c(x,t)$), with diffusion constants $D_m$ and $D_c$ respectively. 
In cells, the diffusion constants of membrane-bound proteins and cytosolic proteins are typically widely different, such that $D_m \ll D_c$. 
The reaction--diffusion equations for the membrane density $m$ and the cytosolic density $c$ read
\begin{subequations} \label{eq:rd-system}
\begin{alignat}{5}
   \label{eq:system_eq1}
   \partial_t m (x,t) 
   &= D_m &&\partial_x^2 \, m 
   &&+ f(m, c),  \\
   \partial_t c (x,t) 
   &= \;D_c &&\partial_x^2 \, c 
   &&- f(m, c),
   \label{eq:system_eq2}
\end{alignat} 	
\end{subequations}
where the reaction term $f(m,c)$ describes the attachment--detachment dynamics of the proteins.
Specific examples of such systems exhibiting self-organized pattern formation can be found in \cite{Ishihara.etal2007,Otsuji.etal2007,Chiou.etal2018,Mori.etal2008}.
At the boundaries, we impose no-flux conditions $ D_c \partial_x c|_{0,L} = D_m \partial_x m|_{0,L}  = 0$. 
The dynamics conserves total protein density
\begin{equation}
	 \bar{n} = \frac{1}{L} \int_0^L \! \mathrm{d}x \; n (x,t),
\end{equation}
with the local total density $n(x,t)= m(x,t) + c(x,t)$. 

To characterize the dynamics and steady states of McRD systems, we recently introduced a framework, termed \emph{local equilibria theory} \cite{Halatek.Frey2018,Brauns.etal2018}.
This theory proposes to analyze spatially extended systems as a collection of small diffusively coupled compartments. 
The local reaction kinetics inside each of the compartments, then serves a proxy for the spatially extended dynamics, enabling a quantitative phase portrait analysis of the spatially extended system in the phase space of reaction kinetics \cite{Brauns.etal2018}.
In the following we briefly review the key results of local equilibria theory for the two-component McRD system and generalize this framework to analyze pattern formation in the presence of a spatial template. 
For a comprehensive analysis of the two-component McRD system on a homogeneous domain, we refer to Ref.~\cite{Brauns.etal2018}.

The reaction kinetics of McRD systems conserves total protein mass, which implies that the reactive flow must point along the reactive phase spaces $n=c+m$, indicated by the gray lines in \fig\ref{fig:no_template_pattern_overview}(a).
The reactive flow vanishes along the \emph{reactive nullcline} (NC), given by $f(m,c) =0$. 
Intersections of the reactive nullcline with reactive phase spaces, given by the total density (mass) $n$, determine the reactive equilibria $(m^*(n),c^*(n))$ shown as black dots in \fig\ref{fig:no_template_pattern_overview}(a). 
Hence, the shape of the nullcline encodes how the reactive equilibria move when total density $n$ is changed, highlighting that the total density $n$ is a \emph{control parameter} for the reaction dynamics. 
Within each reactive phase space, the flow is directed towards a stable reactive equilibrium, as illustrated by the red arrows in \fig\ref{fig:no_template_pattern_overview}(a). 

In a spatially extended system, the total density $n(x,t) = m(x,t) + c(x,t)$ is generically inhomogeneous, and its dynamics is driven by diffusion, as can be seen by adding  Eqs.~\eqref{eq:system_eq1} and~\eqref{eq:system_eq2}
\begin{equation} \label{eq:mass-redistribution}
	\partial_t n(x,t) = D_c \partial_x^2 \eta(x,t),
\end{equation}
where we introduced the \emph{mass-redistribution potential}, defined as \cite{Brauns.etal2018}
\begin{equation} \label{eq:eta-def}
	\eta(x,t) := c(x,t) + \frac{D_m}{D_c} m(x,t).
\end{equation}
To study the interplay of local reactions and diffusive mass-transport in spatially extended systems, local equilibria theory proposes to analyze such systems as a collection of diffusively coupled compartments. 
These notional compartments are chosen small enough that each of them can be regarded as well-mixed.
Thus, local dynamics within each compartment can be characterized in the ODE phase space of reactions which is determined by the density $n(x,t)$ within that compartment.
In this characterization, the \emph{local (reactive) equilibria} and their stability in each local phase space serve as proxies for the reactive dynamics in each compartment.
This becomes clear when one imagines the compartments as isolated, for a given total density profile $n(x)$. 
Then each compartment will approach a stable local equilibrium, parametrized by the local density $n(x)$.
In the spatially coupled system, the total density $n(x,t)$ is diffusively redistributed due to concentration gradients between the compartments (cf.\ Eq.~\eqref{eq:mass-redistribution}).
Consequently, the local equilibria shift  and their stability may change \cite{Halatek.etal2018, Brauns.etal2018}.
This interplay between shifting local equilibria and mass transport is at the core of \emph{local equilibria theory}.

In the remainder of this section, we recapitulate two key results from the phase-portrait analysis of two-component McRD systems \cite{Brauns.etal2018}. We will later generalize this analysis to systems on a spatially heterogeneous domain.

\textit{Flux-balance construction.}\;---\;From the dynamics of the total mass Eq. \eqref{eq:mass-redistribution}--\eqref{eq:eta-def} it follows that, for any stationary pattern (denoted by $\widetilde{m}(x)$, $\widetilde{c}(x)$), $\widetilde{\eta}(x)$ must be constant in space on a domain with no-flux (or periodic) boundary conditions \cite{Brauns.etal2018}: 
\begin{equation} \label{eq:FBS}
	\eta_0 = \widetilde{c}(x) + \frac{D_m}{D_c} \widetilde{m}(x) = \mathrm{const}.
\end{equation}
This relationship defines a linear subspace, termed \emph{flux-balance subspace} (FBS), of the $(m,c)$-phase space of reaction kinetics (purple dashed line in \fig\ref{fig:no_template_pattern_overview}(c)). 
Any stationary pattern must be embedded in a single FBS. This reflects that, in steady state, the diffusive fluxes in $m$ and $c$ are balanced against each other such that there is no net transport of mass.

We can use this condition, Eq. \eqref{eq:FBS}, to geometrically construct the  steady state density profile in the $(m,c)$-phase space and from that estimate the real space density profile. 
The key insight is that we can approximate the concentrations at the plateaus, and the inflection point of the pattern by the local equilibria at the FBS-NC intersections (see \fig\ref{fig:no_template_pattern_overview}(b)). 
We denote these intersection points by $m^-$, $m^0$ and $m^+$, where $m^{\pm}$ correspond to the concentrations at the plateaus  and $m^0$ to the concentration at the inflection point of the pattern (\fig\ref{fig:no_template_pattern_overview}(b,c)). 
Thus, the FBS-offset, $\eta_0$, fully determines these concentrations. 

To determine the FBS-offset, $\eta_0$, one uses that in steady state the net reactive flow within the whole system must be balanced\footnote{Note that before integration, Eq.~\eqref{eq:system_eq1} is multiplied with $\partial_x \widetilde{m}(x)$  as a mathematical trick to substitute the integral over space by an integral over $m$.}:
\begin{equation} \label{eq:total-turnover-balance}
		\int_{m^-(\eta_0)}^{m^+(\eta_0)} \mathrm{d}m \, f\left(m,\eta_0 - \frac{D_m}{D_c}m\right) = 0, 
\end{equation}
where the plateau concentrations far away from the interface are approximated by the FBS-NC intersections $m^\pm(\eta_0)$. 
This \emph{total turnover balance} condition implicitly determines the FBS-offset $\eta_0$. Note that on a large domain (much larger than the interface width, where the approximation $m(0,L) \approx m_\pm$ holds), total turnover Eq.~\eqref{eq:total-turnover-balance}, and hence $\eta_0$, depends only on the function $f$ and the ratio of the diffusion constants.
This implies that $\eta_0$ is not dependent on the average mass $\bar{n}$ in this approximation.

We will show next that the average mass $\bar{n}$ determines the relative size of the low- and high-density regions and with that the position of the pattern's interface.
This interface is marked by the position of the inflection point $x_0$ of the pattern profile.
For a domain size much larger than the interface width, we can neglect the finite width of the interface region, such that the average mass can be approximated by
\begin{equation} \label{eq:mesa-average-mass}
	L \, \bar{n} \approx x_0 \, n^-(\eta_0) + (L-x_0) \, n^+(\eta_0).
\end{equation}
Conversely, $x_0$ can be determined for a given $\bar{n}$.
Thus, this geometric construction, termed \emph{flux-balance construction}, shows that significant features of the steady state profile are determined by the shape of the nullcline. 

\textit{Mass-redistribution instability.}\;---\;In addition to the construction of stationary patterns, it was shown in Ref.~\cite{Brauns.etal2018} that the nullcline shape determines the stability of a homogeneous steady state, and that the mechanism underlying lateral (``Turing'') instability is a mass-redistribution cascade.
Specifically, it was found that a homogenous steady state is laterally unstable when the \emph{slope} of the nullcline $\chi(\bar{n}) := \partial_m c^*|_{\bar{n}}$ is steeper than the slope of the FBS (see Section II.D1 in Ref.~\cite{Brauns.etal2018} for a derivation),

\begin{equation} \label{eq:nc-slope-criterion}
	\chi(\bar{n}) < - \frac{D_m}{D_c},
\end{equation}
which, using the mass-redistribution potential, Eq.~\eqref{eq:eta-def}, is equivalent to $\partial_n \eta^* < 0$.
If this condition is fulfilled, high-density regions act as cytosolic sinks, leading to further accumulation of mass and hence a mass-redistribution cascade. This motivates the corresponding name \emph{mass-redistribution instability}. 

Starting from a homogeneous steady state with a small random perturbation, the initial dynamics is dominated by the fastest growing eigenfunction of the linearized dynamics. 
At the onset of this instability there is a dominant eigenfunction that determines the initial dynamics of the system. 
We can find this dominant eigenfunction by linearizing the system around its homogenous steady state (linear stability analysis). 
For a homogenous steady state, these eigenfunction are Fourier modes and their growth rates are given by the dispersion relation, shown as inset in \fig\ref{fig:no_template_pattern_overview}(d). 
The fastest growing mode, $q_\text{c}$, determines the length scale of the initially growing pattern as illustrated in \fig\ref{fig:no_template_pattern_overview}(d)).
Subsequently the pattern coarsens into a single peak \cite{Ishihara.etal2007,Otsuji.etal2007,Chiou.etal2018,Brauns.etal2018}.

\subsection{Pattern formation with a step-like template}
\label{sec:intro-model}

A common feature of the biological examples we discussed in Section~\ref{sec:motivation} is that the templates have a sharp edge, and that a downstream protein pattern localizes to this edge. 
To obtain a conceptual understanding of how such an edge-sensing mechanism might work, we study how an idealized step-like template affects the pattern formation of the two-component McRD model as a paradigmatic example. 

\begin{figure}
\includegraphics{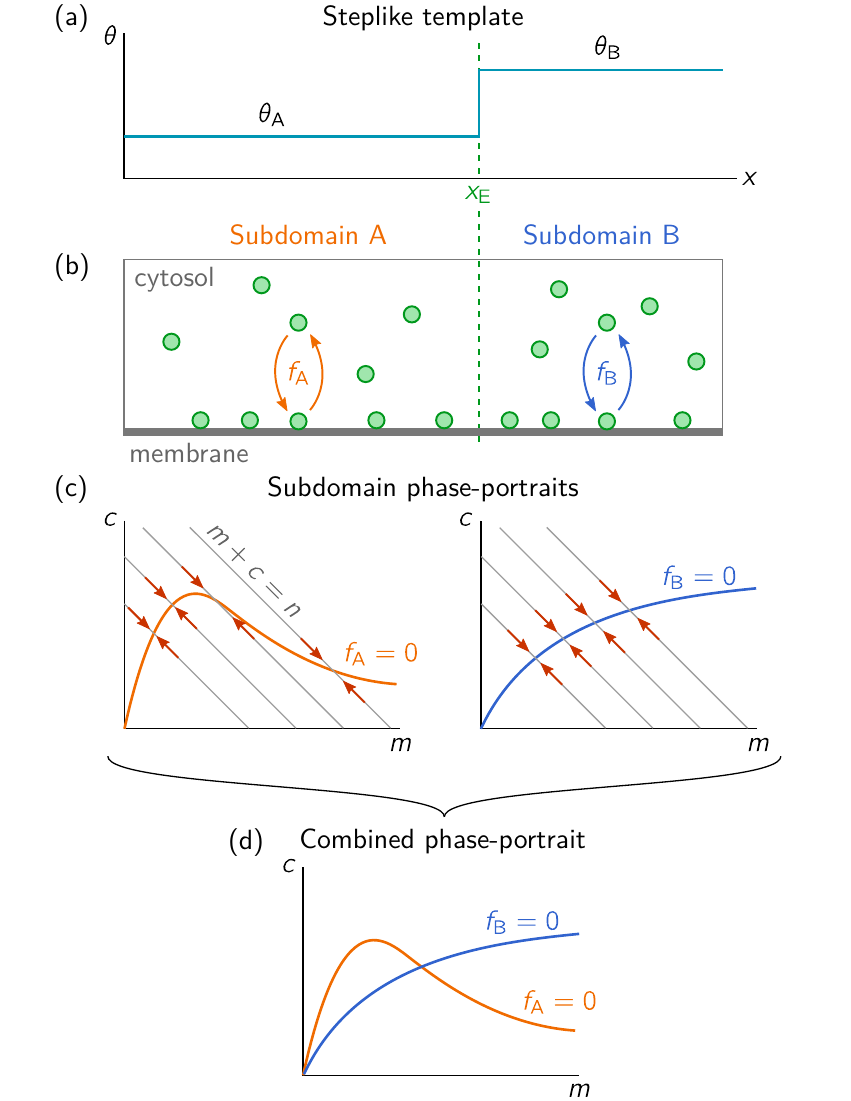}
\caption{\label{fig:template} 
Illustration of a step-like template (a) that acts on the reaction kinetics defines two subdomains, labelled A and B, with respective reaction kinetics $f_\mathrm{A}$ and $f_\mathrm{B}$.
(b) We consider a two-component system describing, for example, a single protein species that cycles between a membrane bound and a cytosolic state.
(c) The reactive flow due to the reaction kinetics can be visualized in the $(m,c)$-phase space of concentrations. Due to mass-conservation, the flow points along the reactive phase spaces $m + c = n$ (indicated by gray lines). Along the nullclines $f_\mathrm{A,B}$, the flow vanishes. Therefore, the reactive flow in each subdomain is qualitatively captured by the shape of the respective nullcline.
(d) With the reactive flow encoded by the nullclines, the phase portraits of the two subdomains can be combined (`overlaid') into a single phase space. This combined phase space will be used for the construction of stationary states.
}
\end{figure}

We consider a step-like template profile $\theta(x)$ with a sharp edge at $\xE$
\begin{align}
 \temp(x) &:=
  \begin{cases}
   \temp_{\mathrm{A}}                 & x \leq \xE\\
   \temp_{\mathrm{B}}                 & x > \xE
  \end{cases} \label{eq:template}\; ,
\end{align}
as illustrated in \fig\ref{fig:template}(a).
Such a template defines two spatial subdomains (labelled A and B).
Here, we consider a template that couples to the downstream pattern forming system via the local reactions $f(m,c; \theta(x)$, such that different reactive dynamics
\begin{equation}
	f_\mathrm{A,B}(m,c) := f(m,c;\theta_{\mathrm{A,B}}),
\end{equation}
govern the system in the two subdomains (see \fig\ref{fig:template}(b)). 

Pattern forming systems with a step-like (also called ``jump-type'') heterogeneity have a rich history in the mathematical literature (see e.g.\ \cite{Maini.etal1992,Ermentrout.Rinzel1996,Bar.etal1996,Yuan.etal2007,Nishiura.etal2007,Miyazaki.Kinoshita2007,Nishi.etal2013,Doelman.etal2016,vanHeijster.etal2019,Scheel.Weinburd2018}) where, they have been studied in the context of front-pinning \cite{Doelman.etal2016}, pulse localization \cite{vanHeijster.etal2019}, and wavenumber selection \cite{Scheel.Weinburd2018}, to name a few recent examples.
These studies predominantly focussed on excitable media and the models studied are not mass-conserving. Furthermore, the prevalent methods employed in these studies are singular perturbation theory (Refs.~\cite{Jones1995,Ward2006} may serve as general introductions) and normal form theory  (see e.g.\ Ref.~\cite{Guckenheimer.Holmes1983}).
The former method uses matched asymptotics and is based on a separation of spatial scales; the latter applies in the vicinity of a bifurcation.

Here, we choose a conceptually different approach building on the recently developed local equilibria theory \cite{Halatek.Frey2018,Brauns.etal2018}, specifically the phase-portrait analysis outlined in Sec.~\ref{sec:recap} and introduced in detail in \cite{Brauns.etal2018}. 
Our starting point is to use the reactive nullclines of the two subdomains as proxies for the respective reactive flows (\fig\ref{fig:template}(c)). 
For this purpose, we combine the phase portraits of the subdomains into a single phase portrait as shown in \fig\ref{fig:template}(d). 
This `overlaying' of the phase portraits facilitates a geometric analysis of the system with a step-like template based on the approach presented previously for the two-component system on a homogeneous domain \cite{Brauns.etal2018} (see recap in Sec.~\ref{sec:recap}).
Throughout the paper we will use nullcline shapes as illustrated in \fig\ref{fig:template}(d) (see Appendix \ref{app:model} for the specific equations and parameters). 
A different nullcline `arrangement', and the general role of the nullcline shapes are discussed in Appendix \ref{app:diff_nullcline}.

The remainder of the paper is structured as follows.
In Sec. \ref{sec:base_state}, we first discuss how the spatial template leads to steady states with a monotonic density profile, which we call \emph{spatially heterogeneous base states}.
 We analyze these base states in the phase space of reaction kinetics and show that they are determined by a balance of reactive turnovers. Importantly, we find that this balance can break down, such that there are regimes where no base state (i.e.\ monotonic steady state) exists.
In Sec.~\ref{sec:stationary} we show that, in these regimes, \emph{stationary patterns} (i.e.\ non-monotonic steady states) emerge, with a density peak either at the system boundary or at the template edge.
Specifically, we show that the peak at the template edge exist if the nullclines intersect at a point where only one of them has negative slope, as illustrated in \fig\ref{fig:template}(d). In Sec.~\ref{sec:regional-instability}, we introduce the concept of \emph{regional instability} to understand the transition between the monotonic base states to the non-monotonic patterns and the conditions under which a peak emerges at the template edge (`edge sensing'). 
Importantly, we find simple geometric criterion for edge sensing: The reactive nullclines of the two subdomains have to intersect at a point where only one of them has negative slope.
Finally, in Sec.~\ref{sec:moving-template}, we ask how the peak responds to a moving template edge and show that this can lead to depinning or suppression of the peak when the template edge moves too fast.

\section{Construction of steady states and their bifurcations}
\label{sec:steady-states}

The goal of this section is to characterize the steady states of the two-component McRD system with a step-like template.
These systems generically don't exhibit spatially homogenous states.
Non-homogeneous steady states can be categorized based on the monotonicity of their density profiles.
In Sec.~\ref{sec:base_state}, we characterize the monotonic steady states, which consist of two plateaus connected by a monotonic interface at the template edge $\xE$, as shown in Fig.~\ref{fig:base_state}(a).
We call these monotonic steady states the \emph{(spatially) heterogenous base-states}. 
In subsection~B we extend our analysis to non-monotonic steady states which are the `genuine' self-organized patterns of the system.

Note that we restrict our construction to the nullcline shapes shown in Fig.~\ref{fig:template}(d) here. Generalizations to other arrangements follow the same principles and can be worked out analogously.

\subsection{Monotonic steady states (base states)}
\label{sec:base_state} 

\begin{figure*}
\includegraphics{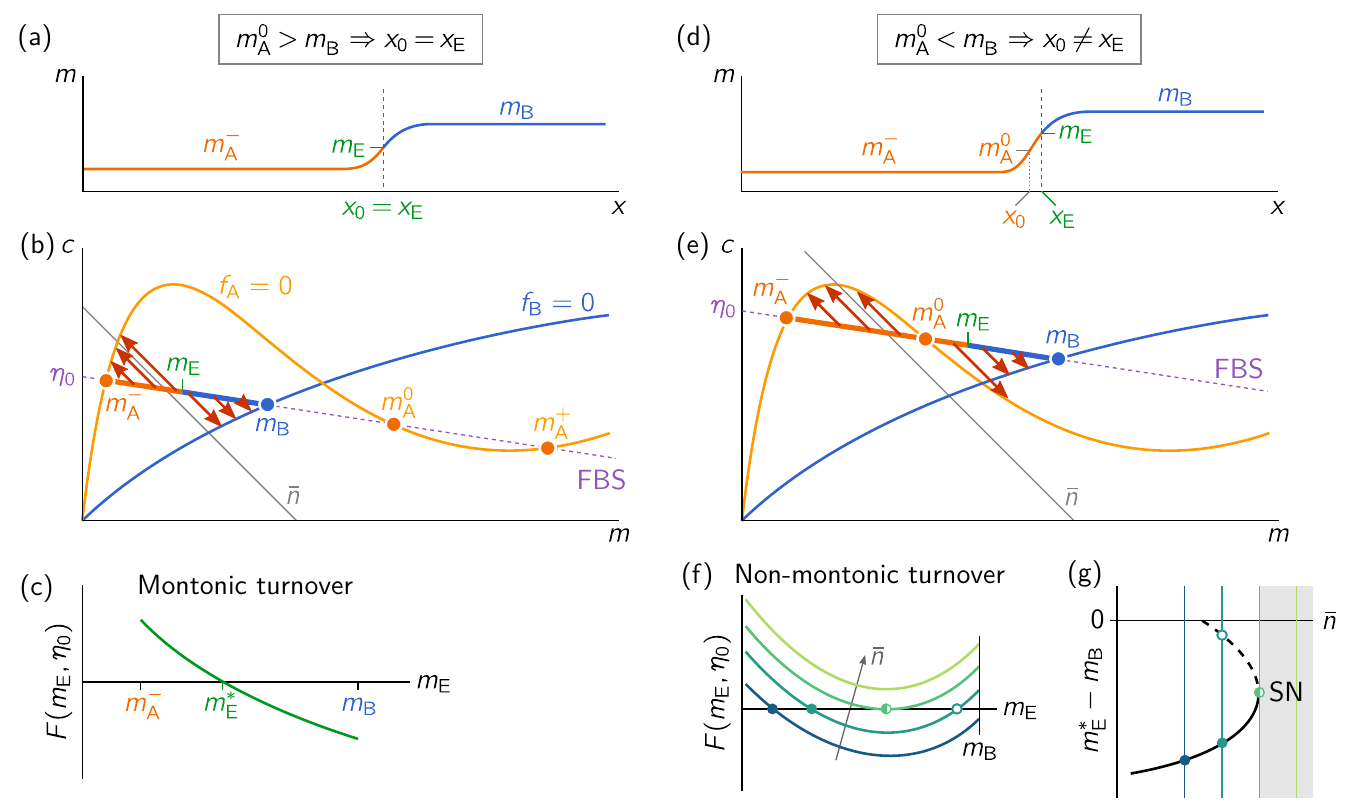}%
\caption{\label{fig:base_state}
Illustration of the construction of monotonic steady states (\textit{heterogeneous base states}) in $(m,c)$-phase space.
(a) Density profile with the inflection point at the template edge $\xE$. 
(b) Phase space including the reactive nullclines of subdomain A (orange) and B (blue) with the corresponding density distribution (thick orange and blue line).
(c) Total turnover as a function of the membrane density at the template edge $\mE$. Total turnover balance determines the steady state value $\mEeq$.
(d) Density profile for a case where the inflection point of the profile, $x_0$, lies within subdomain A, that is, when $m_\mathrm{A}^0 > \mB$ as illustrated in the corresponding phase space representation (e).
(f) In this case, the total turnover becomes a non-monotonic function of $\mE$, such that total turnover balance may have multiple solutions, or no solution at all. These different cases are sketched here for different average global total densities $\bar{n}$. 
(g) Bifurcation diagram of the base state, showing the saddle-node bifurcation due to breakdown of total-turnover balance. In the region beyond the saddle-node bifurcation (shaded in gray) no monotonic base state exists. Note that monotonicity enforces $\mEeq - \mB < 0$ here.
}
\end{figure*}

In steady state, the net diffusive flux that redistributes mass must vanish and reactive flows in $m$ and $c$ must be balanced.
This balance is encoded in the constraint that a stationary pattern's phase space distribution must be embedded in a flux-balance subspace (cf.\ Eq.~\eqref{eq:FBS}). 
This constraint is independent of the local reactions, and, hence, also holds when the local reaction are heterogeneous due to a template (purple dotted line in \fig\ref{fig:base_state}(b)). 

Analogously to the construction of mesa patterns in Sec.~\ref{sec:recap}, we can graphically construct the steady-state density profile in real space, as illustrated in \fig\ref{fig:base_state}(b, e). To that end, we approximate the density profile at the plateaus---where diffusive fluxes cancel everywhere---by the concentration at the FBS-NC intersections, such that the concentration at the plateaus is fully determined by the FBS-offset, $\eta_0$.
Let us denote the membrane concentration at the FBS-NC intersections as $\mA^-(\eta_0)$, $\mA^0(\eta_0)$ and $\mA^+(\eta_0)$ for subdomain A (orange nullcline) and as $\mB(\eta_0)$ for subdomain B (blue nullcline). 
For specificity, we consider in the following a base state that approaches the plateaus $\mA^-$ and $\mB$ far away from the template edge.
An analogous construction can be made for a monotonic state that connects $\mA^+$ and $\mB$.\footnote{
Furthermore, note that we ignore potential intersection points of FBS and B-nullcline at higher masses. 
In this regime, subdomain B will also exhibit lateral instability. 
Here, we restrict ourselves to the regime where only subdomain A becomes laterally unstable.
}

The conservation of average total density $\bar{n}$ enforces a constraint on the construction of the base state.
This constraint can be used to estimate $\eta_0$ from the average total density $\bar{n}$. 
For a domain much larger than the profile interface, the interface region can be neglected and the average total density can be approximated by the weighted average of the plateau densities in the two subdomains
\begin{equation} \label{eq:average_n}
\begin{split}
	& \bar{n} \approx  \xE \, n^-_{\mathrm{A}}(\eta_0) + (L - \xE) \, n^{}_\mathrm{B}(\eta_0),
 \end{split}
 \end{equation}
with $n^-_\mathrm{A}$ and $n^{}_\mathrm{B}$ the total density at the plateaus in subsystem A and B respectively.
This determines an implicit, approximate relation between the control parameter $\bar{n}$ and the FBS-offset $\eta_0$.

Upon changing $\bar{n}$, the plateau concentrations of the density profile must change, and hence, $\eta_0$ must shift (cf. Eq. \eqref{eq:average_n}).
For the laterally stable plateaus, the nullcline slope at the corresponding FBS-NC intersections ($\mA^-$ and $\mB$) is larger than the FBS-slope ($\partial_n \eta^*_\mathrm{A,B}(n) > 0$, cf. NC-slope criterion Eq.~\eqref{eq:nc-slope-criterion}). 
Hence, the relationship $\eta_0(\bar{n})$ must be monotonically increasing for stable base states, as one sees by taking the derivative of Eq.~\eqref{eq:average_n} w.r.t $\eta_0$, and using that monotonicity of a function implies monotonicity of its inverse.

Note that, even though the base state looks similar to a mesa pattern in a system on a homogeneous domain, the relationship between $\eta_0$ and $\bar{n}$ makes a key difference between the two cases. 
As we discussed in the introduction (Sec.~\ref{sec:intro}B), in a system on a homogeneous domain, changing $\bar{n}$ does not affect $\eta_0$ in a system much larger than the interface width, but instead shifts the pattern interface (cf. Eq.~\eqref{eq:mesa-average-mass}).
In contrast, the interface position of a heterogeneous base state is determined by the position of the template edge $\xE$. Hence, to accommodate a given average mass $\bar{n}$ the plateau concentrations $n^-_\mathrm{A}(\eta_0)$ and $n^{}_\mathrm{B}(\eta_0)$, determined by FBS-NC intersection points, must adapt (cf.\ Eq.~\eqref{eq:average_n}). Thus, $\eta_0$ of a heterogeneous base state depends directly on $\bar{n}$.
 
So far we have estimated the concentration at the two plateaus of the monotonic steady state profile. To determine how these two plateaus are connected at the template edge position $\xE$, we use the condition that in steady state the total reactive turnover in the system must vanish. 
In the vicinity of the template edge at $\xE$, the concentrations deviate from the local equilibria, such that there are reactive flows (illustrated by red arrows in Fig.~\ref{fig:base_state}(b)).
Since the template introduces two subdomains with different reaction kinetics, the total reactive turnover in a system with a template is given by the sum over the turnover in the two subdomains,
\begin{equation}  \label{eq:total-turnover} 
\begin{split}
	 F(\mE; \eta_0) &= F_\mathrm{A}(\mE; \eta_0) + F_\mathrm{B}(\mE; \eta_0)\\
	 & = \int_{m^-_{\mathrm{A}}}^{\mE}{dm}\, f_{\mathrm{A}}(m, \eta_0 - \frac{D_m}{D_c} m) \\
	 & + \int_{\mE}^{m_{\mathrm{B}}}{dm}\, f_{\mathrm{B}}(m, \eta_0 - \frac{D_m}{D_c} m),
\end{split}	
\end{equation}
where $\mE$ is the membrane concentration at the template edge.
In steady state, the total turnover $F(\mE; \eta_0)$ must vanish such that all reactive flows in the system balance. 
Thus, the solution of $F(\mEeq; \eta_0) = 0$ (see \fig\ref{fig:base_state}(c)) determines the steady state concentration at the template edge $\widetilde{m}(\xE) = \mEeq$.
Note that, due to monotonicity, $\eta_0$ and $\mEeq$ uniquely identify a base state of a given system.

For small enough $\bar{n}$, the second FBS-NC intersection $\mA^0$ for the A-nullcline is larger than the FBS-NC intersection for the B-nullcline $\mB$ as illustrated in \fig\ref{fig:base_state}(b). 
In this case, both summands of Eq. \eqref{eq:total-turnover} are monotonic in $\mE \in [\mA^-, \mB]$ because the reactive flow does not change sign within either subdomain, i.e. the inflection point of the profile coincides with the template edge.
Hence, there is only a single solution $\mEeq$ that fulfills total turnover balance. 
For larger $\bar{n}$ (and thus $\eta_0$, cf.\ Eq.~\eqref{eq:average_n}), $\mA^0$ can become smaller than $\mB$, as illustrated in the sketch in \fig\ref{fig:base_state}(d,e). 
This entails that the position where the reactive flows change sign (i.e.\ inflection point) lies in subdomain A (see \fig\ref{fig:base_state}(d,e)).
Thus, $F_\mathrm{A}(\mE; \eta_0)$, and thereby also the total turnover $F$ as a function of $\mE$ becomes non-monotonic  and may thus have multiple roots.
Indeed, for increasing $\bar{n}$, our flux-balance construction predicts three different regimes: (i) A regime where there is one solution in the interval $[\mA^-, \mB]$, (ii) a regime with two solutions, and (iii) a regime with no solution (as illustrated in the sketch in \fig\ref{fig:base_state}(f)).
In the last regime, total turnover balance becomes impossible for a \emph{monotonic} steady state (base state).
In Sec.~\ref{sec:stationary}, we will see how total turnover balance can be reached in this regime by a \emph{non-monotonic} steady state.

The roots of $F$ correspond to different base states which we characterize by the amplitude of the density profile in Subdomain~B, $\mEeq - \mB$; see \fig\ref{fig:base_state}(g). 
For monotonic states (i.e.\ base states), $\mEeq - \mB$ is negative\footnote{Note that for high-mass base states, monotonicity enforces $\mEeq - \mB > 0$, in the case of a nullcline arrangement as shown in Fig.~\ref{fig:template}.}.
At the transition from regime~(ii) to~(iii), the base state undergoes a saddle-node bifurcation at $\bar{n}_\mathrm{SN}$.
From the flux-balance construction and total turnover balance, we can estimate the position of this bifurcation.
At the saddle-node bifurcation point, the minimum of $F$ coincides with the root of $F$.
From Eq.~\eqref{eq:total-turnover} it follows that $F$ reaches its minimum at $\fA(m_\mathrm{min}, \eta) = \fB(m_\mathrm{min}, \eta)$.
Thus, this condition, together with $F(m_\text{min};\eta_\text{SN})=0$ implicitly determines the value of $\eta_\text{SN}$ at the saddle-node bifurcation.
From this we can then estimate $\bar{n}^{}_\text{SN}$ via \eqref{eq:average_n}.

To test this approximate construction of steady states, we use specific reaction terms $\fA$ and $\fB$ as specified in Appendix~\ref{app:model} and compare the steady state profiles obtained from the flux-construction to the profiles obtained from numerical continuation (see Appendix~\ref{app:continuation} for a short description of numerical continuation and the comparison of steady states. A more detailed explanation of continuation methods can be found in Ref.~\cite{Krauskopf.etal2007}).
We find that the flux-balance construction gives a estimate of the steady states profiles for sufficiently large system sizes (see Appendix~\ref{app:continuation}).

As we noted above, there is also a family of base states which connects a plateau at $m_A^+$ (instead of $m_A^-$) in subdomain~A to $\mB$ in subdomain~B.
These base states have a high average mass, and we will refer to them as `high-mass' base states. 
Following the same arguments as above, we find that these base states undergo a saddle-node bifurcation when the average total mass is decreased below a critical average mass, analogously to the saddle-node bifurcation of `low-mass' base states discussed in this section. 

In summary, we have shown how to find monotonic steady states (base states) with a flux-balance construction.
Notably, we found that for a range of total mass $\bar{n}$, this flux-balance construction has no solution,
and hence, there exist no monotonic steady states. In this regime, the steady states must be non-monotonic.
We next study these non-monontonic steady states, which we refer to as \emph{patterns}.

\subsection{Non-monotonic steady states (patterns)}
\label{sec:stationary}

\begin{figure*}
\includegraphics[width=1.\textwidth]{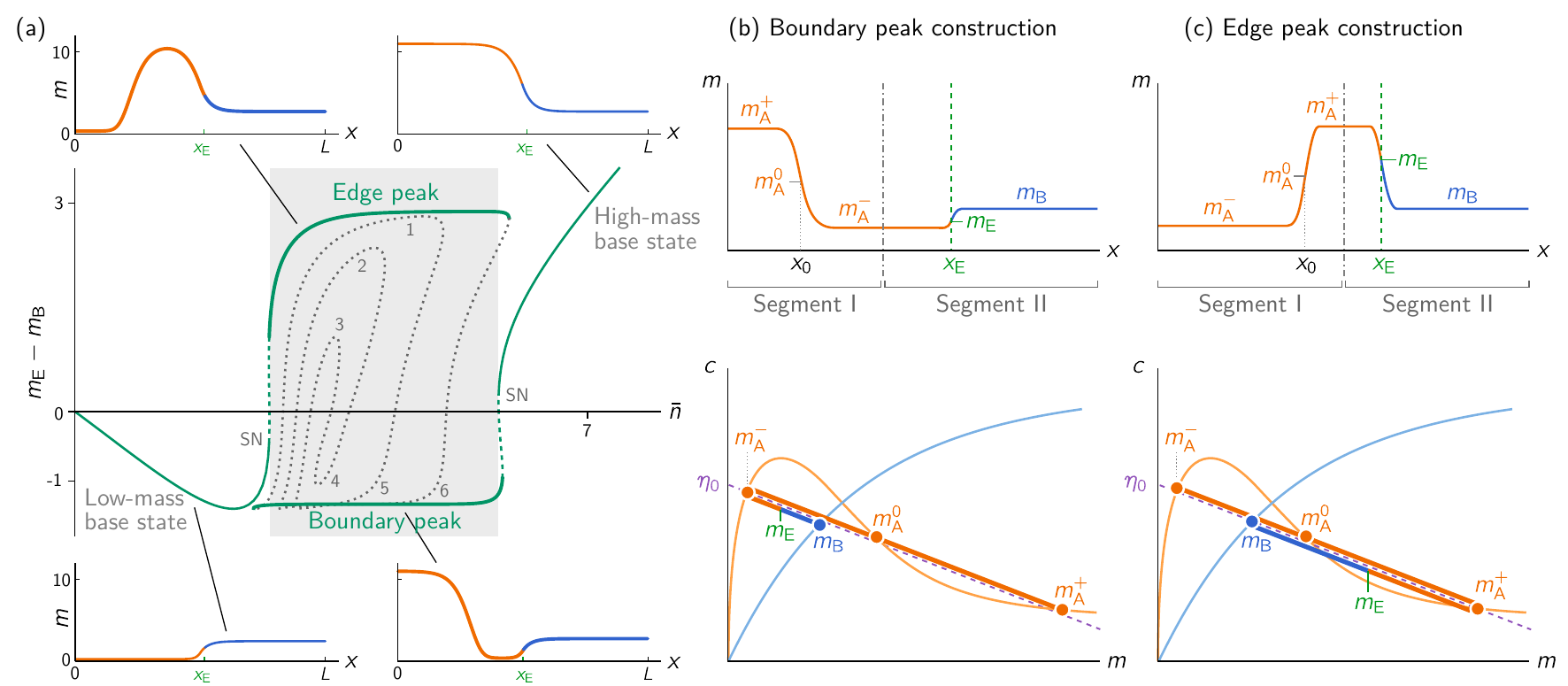}
\caption{\label{fig:stationary-patterns} 
Bifurcation structure and phase space construction of stationary patterns.
(a) Bifurcation structure of stationary states for the average mass, $\bar{n}$, obtained by numerical continuation, together with spatial profiles for the stable steady states (montonic base states and non-monotonic patterns). In the area shaded in gray, no monotonic base states exist. Solid (dashed) lines indicate stable (unstable) branches. The unstable branches in the central region, marked by numbers correspond to unstable patterns with multiple inflection points in subdomain A (see \fig\ref{fig:unstable-states} in Appendix~\ref{app:continuation}). In Supplementary Movies 1 and 2 we show finite element simulations in which we adiabatically increase and decrease the mass through the low- and high-mass saddle-node bifurcations.
(b,c) Sketches of stable, non-monotonic, stationary patterns together with the corresponding phase space constructions. At the density profile's extremum, marked by the dash dotted vertical line, there are no gradients ($\partial_x \widetilde{m} = 0 = \partial_x \widetilde{c}$), such that a notional no-flux boundary can be introduced. Thus, the resulting two segments (labelled I and II), can now be studied separately. In the phase portraits, the density distributions of the two segments connect at the extremal concentrations $\mA^-$ and $\mA^+$, in (a) and (b) respectively. They are shown slightly offset from the FBS (dashed purple line) for visual clarity. The true density distribution must of course be embedded in a single FBS to fulfill diffusive flux-balance. \\
Parameters for the bifurcation diagram:
  $ D_m = 0.01 $, 
  $ D_c = 0.5 $, 
  $ \hat{k}_{\mathrm{fb}} = 0.25 $, 
  $ \hat{k}_{\mathrm{off}} = 4 $, 
  $ \temp_{\mathrm{B}} = 20 $, 
  $ \temp_{\mathrm{A}} = 2 $, 
  $ L = 10 $,  
  $ \xE = 5 $.
}
\end{figure*}

To gain some intuition about the structure of the stationary patterns, we first calculate them numerically as a function of the average total density $\bar{n}$ using numerical continuation\footnote{See Appendix~\ref{app:continuation} for a brief description of the core idea behind numerical continuation. An excellent overview is provided in Ref.~\cite{Krauskopf.etal2007}.} for a specific choice of the reaction term $f(m,c;\theta)$ specified in Appendix~\ref{app:model}.
The resulting one-parameter bifurcation structure shows that the system exhibits two stable patterns, one with a peak (high density region) at the template edge and one with a peak at the system boundary, respectively (see \fig\ref{fig:stationary-patterns}(a)).
In the bifurcation structure, the two branches of stable patterns are connected by a cascade of unstable patterns exhibiting multiple peaks; see \fig\ref{fig:unstable-states} in Appendix~\ref{app:continuation}. 
Their instability can heuristically be understood as a coarsening process due to a competition for total density, similarly as in a system on a homogeneous domain \cite{Brauns.etal2018}. 
Some more technical aspects of this bifurcation structure are discussed in Appendix~\ref{app:continuation}.

Can we use the flux-balance construction to construct non-monotonic steady states as well?
At the extrema of any stationary pattern in a two-component McRD system, the gradients (and hence diffusive flux) in \emph{both} membrane and cytosol concentration vanish simultaneously (cf.\ diffusive flux-balance Eq.~\eqref{eq:FBS}). 
This allows us to place notional reflective boundaries at extrema, effectively splitting the non-monotonic profile into monotonic segments. 
Thus, we can use the flux-balance construction as described in Sec.~\ref{sec:base_state} to construct the steady states in the two segments separately, with the additional constraint of continuity at the boundaries connecting the segments.

The stable patterns in the two-component McRD system with a step-like template have only a single peak, and hence only a single extremum within the domain that splits the system into two segments (labelled I and II; see \fig\ref{fig:stationary-patterns}(b,c)).
Segment~I is fully embedded in subdomain~A, i.e.\ it is a system on a homogeneous (sub)domain.
Hence, for sufficiently large domain size, its steady state is a mesa pattern\footnote{In the vicinity of the saddle-node bifurcations of these patterns, segment~I exhibits a peak pattern instead of a mesa pattern. To obtain an approximation for this case, one would need to generalize the peak approximation as discussed in Ref.~\cite{Brauns.etal2018}.} as introduced in Sec.~\ref{sec:recap}.
The orientation of the mesa pattern in segment~I determines whether the density peak is located at the left domain boundary or at the template edge.
segment~II contains the template edge, such that the steady state in segment~II is a heterogeneous base state.

By continuity, the FBS-offset $\eta_0$ must be identical in both segments.
Recall that for a mesa pattern, $\eta_0$ is determined by total turnover balance, and independent of the average mass and domain size in the large domain size limit (see Sec.~\ref{sec:recap} and Ref.~\cite{Brauns.etal2018}).
We can thus find $\eta_0$ solely by total-turnover balance in segment~I, without specifying the position of the boundary between the segments, and without specifying the average masses in the two segments respectively.
Instead, given $\eta_0$, we find the average mass in segment~II, $\bar{n}_\mathrm{II}$, via Eq. \eqref{eq:average_n}, which depends on the choice for the orientation of the mesa pattern in segment~I.
In segment~II, subdomain~B plays the role of a mass-reservoir that absorbs a fraction of the total average mass and thus reduces the mass available to the mesa pattern in segment~I, $\bar{n}_\mathrm{I} = \bar{n} - \bar{n}_\mathrm{II}$.
Finally, $\bar{n}_\mathrm{I}$ determines the position of the mesa pattern's interface in segment~I via Eq. \eqref{eq:mesa-average-mass}.
Similarly, we can construct the (unstable) patterns with multiple peaks by splitting the system into more than two segments. 

We conclude that the flux-balance construction fully characterizes the stationary patterns of the system with a step-like template. These steady states exhibit density peaks similar to a system on a homogeneous domain, however, the position of the density peak depends on the template edge position.

We next ask which of the two stable patterns emerges as the base state ceases to exist. 
To that end, we use finite element simulations and adiabatically increase the total average density in a system such that it passes through the base-state bifurcation (Supplementary Movie~1).
The system evolves into a pattern with a peak at the template edge (corresponding to the upper branch in \fig\ref{fig:stationary-patterns}(a)).
Upon further increase of $\bar{n}$ the peak widens and eventually transitions into a mesa pattern at the template edge. The right hand interface of the mesa patter remains localized at the template edge while the left hand interface moves into subdomain~A to accommodate the additional mass (cf.\ Eq.~\ref{eq:mesa-average-mass}).
Eventually for even larger $\bar{n}$, the mesa pattern ceases to exist as its interface hits the left boundary of the domain. The system then transitions to the `high-mass' base state which connects the FBS-NC intersection points $\mA^+$ and $\mB$.
Going backwards by adiabatically decreasing $\bar{n}$, the system passes through the `high-mass' base state's saddle-node bifurcation.
The corresponding regional instability leads to the formation of a trough, rather than a peak, at the template edge.
The resulting stationary pattern has a minimum at the template edge and a maximum at the left boundary (corresponding to the lower branch in \fig\ref{fig:stationary-patterns}(a)) (Supplementary Movie~2).
Upon further decrease of $\bar{n}$, the interface of this pattern will reach the left boundary of the domain such that the system transitions back into the low-mass base state. 

These transitions also take place when the average mass is changed non-adiabatically but still so slow that mass-transport across the system by cytosolic diffusion (${\sim} L^2/D_c$) is faster than the rate at which mass is added or removed.
Interestingly, when we increase the mass on a non-adiabatic timescale we observe multiple transient patterns, which we characterize in Appendix \ref{app:non-adiabatic}.
Furthermore, we show that we can get similar transitions between the base state and the patterns if, instead of increasing the average mass, the local reactions $f_\text{A,B}$ in the two subdomains are varied over time (Appendix~\ref{app:temp_template}).

Taken together, we have shown that the flux-balance construction can be used to construct non-monotonic steady states by splitting the density profile into monotonic segments. 
This is possible because, the stationary pattern profile can be split at extrema, where all diffusive fluxes vanish. 
We found that the system can exhibit two patterns, with a density peak either at the system boundary or at the template edge. 
The peak at the template edge only exists when the two nullclines intersect at a point where only one of them has negative slope. 
Furthermore, our finite element simulations show that increasing the mass, starting from the low-mass base state, leads to a peak at the template edge. 
Vice versa, decreasing mass, starting from the high-mass base state leads to a peak at the system boundary. 
In the next Section we provide a heuristic argument to understand under which conditions the peak emerges at the template edge.

\subsection{Template-induced regional instability}
\label{sec:regional-instability} 

We next want to understand the mechanism of pattern selection as the system goes through the saddle-node bifurcation(s) where the base state ceases to exist.
We first show that a (numerical) linear stability analysis explains why either a peak or a trough pattern grows at the template edge, as the system goes through the bifurcation. 
We then provide a heuristic argument to explain this edge-sensing mechanism, and formulate a geometric criterion under which this mechanism works, based on the shape of the nullclines.

To study how the base state develops into a pattern, as the system goes through the saddle-node bifurcation, we consider a base state $(\widetilde{m}(x),\widetilde{c}(x))$ in the vicinity of the bifurcation point and analyze the dynamics of a small perturbation $(\delta m(x,t),\delta c(x,t))$. The dynamics of the perturbed state, up to linear order is given by:
\begin{subequations} \label{eq:linearized-dynamics}
\begin{align}
	\partial_t \delta m(x,t) &= D_m \partial_x^2 \delta m + \tilde{f}_m(x) \delta m + \tilde{f}_c(x) \delta c \, , \\
	\partial_t \delta c(x,t) &= \kern0.3em D_c \partial_x^2 \delta c \kern0.45em - \tilde{f}_m(x) \delta m - \tilde{f}_c(x) \delta c \, . 
\end{align} 
\end{subequations}
The linearized reaction coefficients
\begin{equation}
	\tilde{f}_{m,c}(x) = \partial_{m,c} f \big|_{(\widetilde{m}(x),\widetilde{c}(x))}
\end{equation}
are not homogeneous in space and hence Eq.~\eqref{eq:linearized-dynamics} is a set of linear PDEs with nonconstant coefficients.
We seek solutions of the form $\delta m(x,t) = \Phi_m(x) e^{\sigma t}$, $\delta c(x,t) = \Phi_c(x) e^{\sigma t}$. 
With this ansatz, Eq.~\eqref{eq:linearized-dynamics} turns into the Sturm--Liouville eigenvalue problem
\begin{subequations}
\begin{align}
	\sigma \Phi_m(x) &= D_m \partial_x^2 \Phi_m + \tilde{f}_m(x) \Phi_m + \tilde{f}_c(x) \Phi_c \, , \\
	\sigma \Phi_c(x) &= \kern0.33em D_c \partial_x^2 \Phi_c \kern0.4em - \tilde{f}_m(x) \Phi_m - \tilde{f}_c(x) \Phi_c \, ,
\end{align}
\end{subequations}
for the eigenvalues $\sigma$ and the associated eigenfunctions $(\Phi_m,\Phi_c)(x)$.

The defining feature of a a saddle-node bifurcation is that one eigenvalue vanishes exactly at the bifurcation point.
The associated eigenfunction reveals the flow structure of the dynamics on the slowest timescale close to the bifurcation point (center manifold, see e.g.\ Ref.~\cite{Guckenheimer.Holmes1983}). From this we can gain intuition about the fate of the system upon passing through the bifurcation.

\begin{figure}
\includegraphics{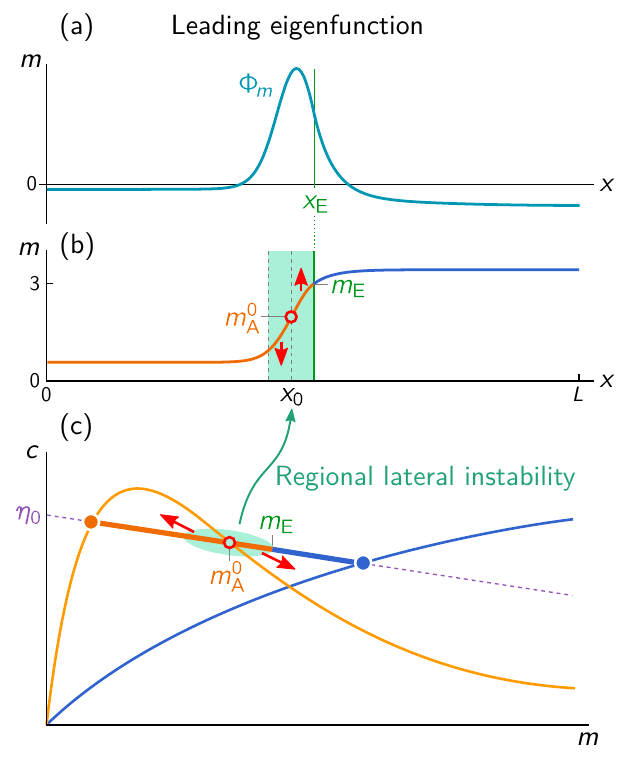}
\caption{\label{fig:regional_instability} 
Regional lateral instability at the base state's saddle-node bifurcation.
(a) The numerically calculated eigenfunction, $\Phi_m$, associated to the vanishing eigenvalue at the saddle-node bifurcation is localized at the template edge. (Parameters as in Fig.~\ref{fig:stationary-patterns} at the saddle-node bifurcation $\bar{n} \approx 2.65541$).
(b) The concentration profile $\widetilde{m}(x)$ of the base state at the saddle-node bifurcation. A spatial region that is fully contained in subdomain A and centered around the profile's inflection point $x_0$ is marked in green.
(c) This spatial region corresponds to a phase space region where the dynamics is guided by a section of the nullcline with a negative slope, i.e.\ where the system is laterally unstable. 
(The phase space is shown as a sketch for visual clarity. See Fig.~\ref{fig:app_role-of-nullcline-shapes}(a) in Appendix~\ref{app:diff_nullcline} for a plot from numerical simulation.)
}
\end{figure}

At the base-state saddle-node bifurcation (marked SN in Fig.~\ref{fig:stationary-patterns}(a)), the numerically calculated\footnote{The Sturm--Liouville eigenvalue problem at the numerically calculated base-state saddle-node bifurcation is solved by discretizing the spatial derivatives (Laplace operator) and solving the resulting eigenvalue problem numerically in Mathematica. For further details see e.g.\ Ref.~\cite{Pryce1993}.} eigenfunction is peaked in the vicinity of the template edge (see \fig\ref{fig:regional_instability}(a)).
This localized eigenfunction indicates that the density profile will change most in the vicinity of the template edge, giving rise to either a peak or a trough as the system goes through the saddle-node bifurcation.

An intuition why the neutral eigenfunction at the base-state saddle-node bifurcation is peaked at the template edge can be gained from the phase portrait as sketched in \fig\ref{fig:regional_instability}(c)).
Recall, that the inflection point of the base state's density profile $\mA^0 = \widetilde{m}(x_0)$ lies within subdomain~A (cf.\ \fig\ref{fig:base_state}(d,e)).
Consider a region centered around $x_0$, fully contained within subdomain A, as indicated in green in \fig\ref{fig:regional_instability}(b).
In phase space this point lies on a section of the A-nullcline with a slope steeper than the FBS.
Suppose for a moment that this region is isolated from the rest of the system.
Then, as the slope of the nullcline at $\mA^0$ is steeper than the slope of the FBS, the homogeneous equilibrium in this region will be unstable due to a mass-redistribution instability.
This instability will set in when the region is large enough to contain the shortest growing mode\footnote{More precisely, the size of the region centered around the inflection point, $2(\xE - x_0$), must be larger than the wavelength of the shortest growing mode $\pi/q_\text{min}(n_0)$.}.
We call this a \textit{regional (mass-redistribution) instability}.

\begin{figure}[b]
\includegraphics{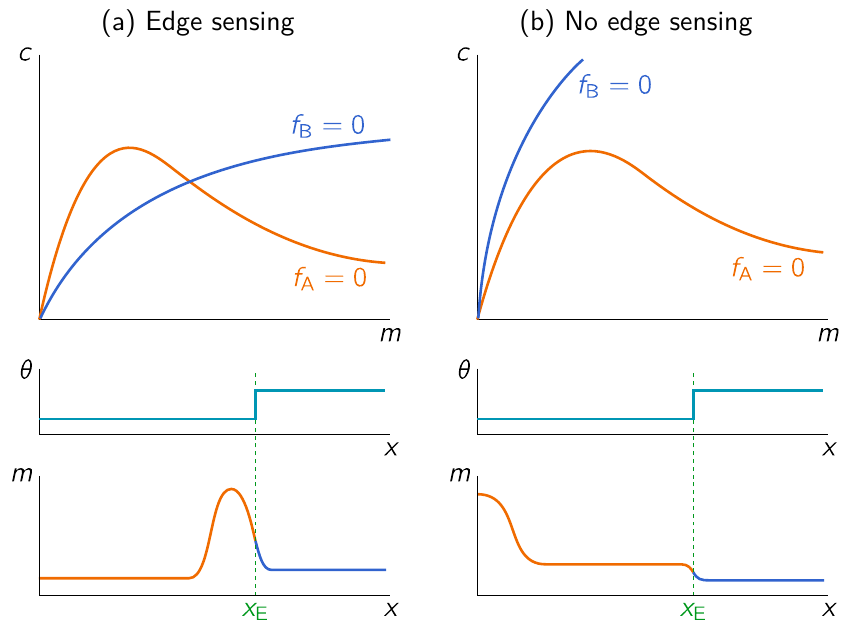}
\caption{\label{fig:nullcline-criterion} 
Nullcline criterion for edge sensing, i.e.\ the emergence of a stable density peak at the template edge. Edge sensing is possible is the nullclines intersect in a point where one of them has negative slope (a). If they don't intersect (b) or intersect in a point where they have both positive (or both negative) slope, stable peaks only exist at a system boundary, and hence, edge sensing is not possible.
}
\end{figure}

A necessary condition to trigger a regional instability at the template edge is that the nullclines of the two subdomains cross at a point where the the A-nullcline fulfills the nullcline-slope criterion  for lateral instability, Eq.~\eqref{eq:nc-slope-criterion} (see Fig.~\ref{fig:nullcline-criterion}(a)).
When this edge-sensing criterion is not fulfilled, as shown in Fig.~\ref{fig:nullcline-criterion}(b), the regional instability sets in at the system boundary first, giving rise to a peak at the system boundary and not at the template edge (as illustrated in Fig.~\ref{fig:app_role-of-nullcline-shapes}(b) in Appendix~\ref{app:diff_nullcline}).
Because the shapes of the nullclines in the two subdomains relative to each other depend on how the template affects the reaction kinetics, the edge-sensing criterion constrains models that can exhibit edge sensing.
In Appendix~\ref{app:cdc42} we demonstrate that the edge-sensing criterion precisely predicts the regime of edge sensing for a phenomenological model of Cdc42. We furthermore show that edge sensing only works if the template increases both the attachment and detachment rate of Cdc42 in one of the subdomains. This may help to identify the relevant molecular players in biological systems.

The concept of regional instability was already discussed in Ref.~\cite{Brauns.etal2018} in the context of excitability (``nucleation and growth'') for a homogeneous domain.  
There, a stable homogeneous steady state is perturbed by moving mass from the system into small region. When this region contains sufficient mass, it can become laterally unstable and thus formation of a peak pattern is ``nucleated.''
With that, the regional instability at the template edge can also be understood in terms of lateral excitability.
From the perspective of subdomain~A, the differing reaction kinetics in subdomain~B induces a perturbation at the subdomain interface, that is, the template edge (orange line in \fig\ref{fig:regional_instability}(b)).
In that sense, the base state is a perturbed homogeneous steady state in each subdomain.
If in subdomain~A, this perturbation becomes large enough to cross the nullcline in a section of negative slope (\fig\ref{fig:regional_instability}(c)), it triggers a lateral instability and thus the formation of density peak at the template edge.
This relationship to excitability highlights that it is the spatial \emph{gradient} of the reaction kinetics due to the template that localizes the instability. 
Heuristically this can be pictured as sensing the spatial derivative of the template.
Here we focussed on a sharp template edge. 
For future work, it will be interesting to study also a smooth template edge. 
Intuitively, if the template gradient is too shallow, it will not induce a laterally unstable region, because the induced deviation from the local equilibria that effectively acts as the perturbation in the analogy to excitability will be too small.

In conclusion, we showed in this section that the template localizes the patterns and determines the position of the instability from which they emerge.
Importantly, this instability determines which of the two stable stationary patterns forms when the system passes through a bifurcation of the base state. 
Finally, we presented a simple geometric criterion, shown in Fig.~\ref{fig:nullcline-criterion}, that determines when the regional instability is localized at the template edge.

\section{Moving template edge}
\label{sec:moving-template}
\begin{figure}
 \includegraphics{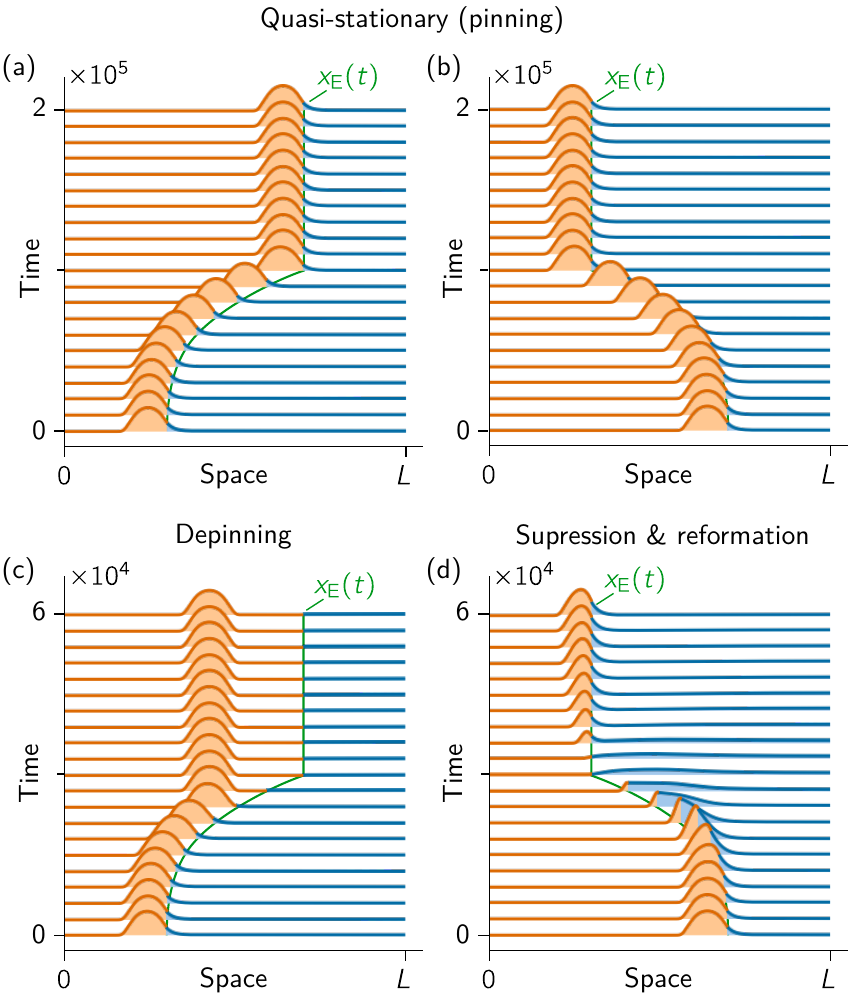}
 \caption{\label{fig:moving_edge} 
 Time evolution of $ m $-profile due to dynamic template interface position. The supplementary material contains a movie for each of the four scenarios.
   (a) Slow pull (the template edge moves away from the peak): This results in pinning of the peak to the template edge (see SI Movie~3). 
   (b) Slow push (the template edge moves towards the peak): This results in pinning of the peak to the template edge (see SI Movie~4).
   (c) Fast pull: This results in depinning as the peak cannot follow the template edge. However, the peak stays where it depins as it is quasi-stable in region A on the observed timescale (see SI Movie~5).
   (d) Fast push: This results in suppression as the peak is not stable in region B. As the peak dissipates the average mass at the new edge position slowly increases and potentially (if there is enough total mass in the system) leads to to a re-entrance of the peak (see SI Movie 6).\\
   Parameters: 
   $ D_m = 0.01 $,  
   $ D_c = 10 $,  
   $ k_{\mathrm{on}} = 1 $,  
   $ \hat{k}_{\mathrm{fb}} = 1 $,  
   $ k_{\mathrm{off}} = 2 $,  
   $ \hat{K}_{\mathrm{d}} = 1 $,  
   $ \bar{n} = 5 $,  
   $ \temp_{\mathrm{B}}= 20 $,  
   $ \temp_{\mathrm{A}} = 0.5 $,    
   $ L = 20 $, 
   $ \xE  = 3/5 \, L $,
   $ v_{\mathrm{E}}(t) = \pm 2.4 \times 10^{-14} \, t^2 $ in (a) and (b),
   $ v_{\mathrm{E}}(t) = \pm 8.9 \times 10^{-13} \, t^2 $ in (c) and (d).
 }
\end{figure}

Until now we considered pattern formation with a fixed template edge position $\xE$.
We next ask what happens when $\xE$ moves after a peak at the template edge has been established (\fig\ref{fig:moving_edge}). 

When the template edge moves, the peak must adapt to the new template edge position. 
In order to reach the new stationary state, mass must be transported from one side of the peak to the other. 
Thus, we expect that the peak follows the template edge position as long as the mass is transported faster than the velocity $v_\mathrm{E}$ at which the template edge moves, i.e. for $v_\mathrm{E} \ll D_c/w$, where $w$ is the width of the peak.
To test the intuition for this adiabatic case, and study what happens in the non-adiabatic case of a fast-moving template edge, we turn to numerical simulations. 
To probe a range of template velocities $v_\mathrm{E}$, we quadratically increase the template edge velocity during the simulation. At a distance $0.3\, L$ from the system boundary, the template movement is stopped.
Furthermore, we move the interface either to the right, away from the peak \fig\ref{fig:moving_edge}(a,c) or to the left, towards the peak \fig\ref{fig:moving_edge}(b,d). 
In the adiabatic case, we find---in agreement with our expectation---that the peak remains pinned to the template edge position (see \fig\ref{fig:moving_edge}(a,b) and SI Movies~3 and~4).

In the non-adiabatic case, when the template edge moves faster than the peak can follow by diffusive mass transfer, the peak position will shift relative to the template edge.
We find that, when the template moves away from the peak (`pulling'), the peak depins at a critical velocity and stops moving (see \fig\ref{fig:moving_edge}(c) and SI Movie~5).
In contrast, when the template moves towards the peak (`pushing'), the peak is suppressed at some critical edge-velocity (see \fig\ref{fig:moving_edge}(d) and SI Movie~6).
Interestingly, the critical velocity for depinning while pulling is much lower than the critical velocity for suppression while pushing.

To heuristically understand this difference of critical velocities, we take a closer look at how the peak adapts to a shifted template position.
Suppose for a moment that we keep the peak profile frozen while shifting the template edge $\xE$ by a small amount $\delta \xE$. This will change the local reactions in the vicinity of the template edge.
When the peak profile is now ``released'', these reactions will lead to changes in the concentrations $m$ and $c$, and hence in the 
mass-redistribution potential $\eta(x,t)$ between the original and shifted edge positions $\xE$ and $\xE + \delta \xE$. The resulting $\eta$ gradient leads to mass transport (Eqs.~\eqref{eq:mass-redistribution}), which in turn causes a movement of the peak.\footnote{A geometric analysis in phase space, similar to the one presented in Figs.~\ref{fig:base_state} and~\ref{fig:stationary-patterns}, shows that the $\eta$ gradient is always such that the peak moves in the direction that the template edge was shifted, until it reaches it's new stationary (pinned) position at the shifted edge position.}
The difference between pulling and pushing, i.e.\ moving the template edge away from the peak or towards it, is the amplitude of the $\eta$ gradient that builds up as the template edge is shifted.
In the case of pulling, the template edge moves into the flat tail of the peak, such that the $\eta$ gradient decreases with increasing distance between peak and template edge. 
In contrast, while pushing, the template edge moves into the steep interface of the peak, leading to an continually increasing $\eta$ gradient.
Only when the template edge has shifted beyond the maximum of the peak, the induced $\eta$ gradient starts decreasing again.

This effect qualitatively explains the different critical velocities for depinning (while pulling) and suppression (while pushing).
In the former case the induced gradients in $\eta$, and hence, the rate of mass transport that shift the peak towards the moving template edge are small and decreasing with peak-to-edge distance. Therefore, depinning is self-enhancing. 
In the case of pushing, the $\eta$ gradient keeps increasing as the peak-to-edge distance decreases. This in turn, increases the speed of the peak due to faster mass transport.
Only when the template edge has crossed the peak maximum, the peak will be suppressed because it then lies mostly within the laterally stable subdomain $B$ that does not support a stable peak.

In summary, we showed that in the case of an adiabatically slow template motion, the peak stays pinned at template edge. 
In the non-adiabatic case we found a qualitatively and quantitatively different behavior between pushing and pulling. Pulling leads to depinning, while pushing eventually leads to suppression.
Furthermore, we heuristically explained why critical velocity for pulling (depinning) is lower than the critical velocity for pushing (suppresion).
Going forward, it would be interesting to study this behavior more systematically, both in numerical simulations and on an analytic level. For example, concepts like response functions \cite{Ziepke.etal2016}, projection onto slow manifolds \cite{Nishiura.etal2007} and singular perturbation theory \cite{Doelman.etal2016,vanHeijster.etal2019} may help to estimate the critical velocities for depinning and suppression.


\section{Discussion}
\label{sec:con}

We showed how protein pattern formation can be controlled by a spatial template (e.g.\ an upstream protein pattern), which acts on the proteins interaction kinetics. 
In particular, we demonstrated for two-component McRD systems how a step-like template---which defines two subdomains with different reaction kinetics---can localize the formation of a peak pattern to the template edge.
We explained this edge-sensing mechanism by a regional (mass-redistribution) instability that emerges at the template edge position. 
This is in contrast to pattern formation on a homogeneous domain, where the instability is generically ``delocalized'' (Fourier modes, cf.\ Fig.~\ref{fig:no_template_pattern_overview}(d)), such that noisy initial conditions have a strong impact on pattern formation process.

Our analysis is based on a recently developed theoretical framework \cite{Halatek.Frey2018,Brauns.etal2018}, termed local equilibria theory. 
This theory proposes to analyze McRD systems as dissected into diffusively coupled compartments, so small that each of the compartments can be considered as well-mixed.
For the paradigmatic (minimal) case of two-component systems, this framework enables one to to perform phase-portrait analysis of the interplay between local reactions and diffusive transport in the phase space of the reaction kinetics.
Here, we have extended this phase-portrait analysis to incorporate the effect of a step-like heterogeneity of the reaction kinetics in the spatial domain. 
We were able to construct the bifurcation diagram for the average mass $\bar{n}$, which is a natural control parameter as it can be controlled by production or degradation/sequestration of proteins in cells (e.g.\ in a cell cycle dependent manner).
We found that, at a critical average mass, the system's base state undergoes a saddle-node bifurcation, such that the system transitions to a stationary pattern, with either a peak at the template edge or at the system boundary. 
The peak forms at the template edge if the template triggers a regional (mass-redistribution) instability at the template edge. 
The phase-portrait analysis enables us to formulate a geometric criterion for this edge-sensing mechanism.
In particular, we show that the step-like template can trigger a regional instability at the template edge, if the nullclines of the reaction kinetics in the two subdomains intersect at a point where one of them has a negative slope (more precisely, a slope steeper than the negative ratio of the diffusion constants, $-D_m/D_c$), as illustrated in Fig.~\ref{fig:nullcline-criterion}. 
Finally, we showed that the edge-localized peak is stable when the template edge is slowly moved and demonstrated that qualitatively different processes---depinning vs.\ suppression---lead to the loss of the edge-localized peak when the template is shifted too rapidly away from the peak (`pulling') or towards it (`pushing').

We speculate that the edge-sensing mechanism studied here might underly formation of the actomysin ring during macropinocytosis and cellular wound healing, as well as the direction of Cdc42 polarization in budding yeast adjacent to the previous bud-site.
In macropinocytosis a high density PIP3 domain has been suggested to act as a template for a ring of actin nucleators that localize to its periphery \cite{Veltman.etal2016}.
Similarly, during cellular wound healing, the inside Abr could act as  a template for the outside Cdc42 ring which then drives recruitment of actin via formins \cite{Vaughan.etal2011}.
Finally, in budding yeast, landmark proteins that localize to the previous bud-site can be pictured as a template that suppresses Cdc42 accumulation at the previous bud-site and simultaneously localizes Cdc42 cluster to its vicinity \cite{Chant.Pringle1995,Tong.etal2007,Lo.etal2013,Miller.etal2017}.
Furthermore, the spatio-temporal organization of intracellular membranes, like the Golgi apparatus, endosomes, and the endoplasmatic reticulum, involves cascades of coupled GTPases \cite{Pfeffer2012,Novick2016,Noack.Jaillais2017}.
Hence, we speculate that this organization may rely on similar domain-edge sensing mechanisms.

The edge-sensing criterion (cf.\ Fig.~\ref{fig:nullcline-criterion}) based on the shape of the nullcline, may provide guidance for the mathematical modeling of these systems and thereby help to identify the key bio-molecular players and processes.
The nullcline shapes of a given model constrain the ability of this model for edge-sensing.
As an example, we showed  for an phenomenological two-component model for Cdc42 pattern formation \cite{Mori.etal2008} that edge sensing requires a template which increases both the attachment and detachment rate of Cdc42 in one subdomain. 
Indeed in single-cell wound healing, the protein Abr could provide such a template for Cdc42 since it is both a guanine exchange factor (GEF)  and a GTPase-activation protein (GAP) for Cdc42 \cite{Vaughan.etal2011}.

Beyond the understanding of living systems, our results may also advance the field of synthetic biology.
Previous studies have explored mechanisms by which a gradient can position a sharp front pattern via a bistable reaction-diffusion system \cite{Rulands.etal2013,Zadorin.etal2017}. 
The edge-sensing mechanism, presented in this paper, is a candidate for a further building block to design spatial protein patterns. 

In future studies, it would be interesting to generalize our results beyond the paradigmatic case of a single, stationary, step-like template in one spatial dimension.
Templates with multiple steps may be dissected into segments with single steps that can then be studied separately. 
Furthermore, it has been shown that the geometry of a cell indirectly affect the attachment--detachment kinetics via the ratio of bulk-volume to surface-area \cite{Thalmeier.etal2016,Ziepke.etal2019}, and curvature sensing proteins \cite{Haupt.Minc2018}.
Another promising direction is to incorporate the dynamics of the template itself as a self-organized pattern-forming system, and include a feedback from the downstream pattern to the template. Such feedback may give rise to complex spatio-temporal behavior like oscillatory patterns and traveling waves that can then be characterized by building on the phase-portrait analysis presented here.
Our results on the moving template (Sec.~\ref{sec:moving-template}) and non-adiabatic upregulation of average mass (Appendix~\ref{app:non-adiabatic}) indicate that the edge sensing works beyond the adiabatic regime.

Finally, even for the elementary case studied here, many important questions remain open.
We showed that a moving template will lead to a loss of the edge-localized peak due to depinning (while pulling) or suppression (while pushing) at different critical edge-velocities. In future work, these transitions should be studied in more detail both numerically and analytically, e.g.\ using a response function formalism \cite{Ziepke.etal2016}.
Furthermore, an analytic approach employing asymptotic methods like singular perturbation theory \cite{Ward2006,vanHeijster.etal2019,Doelman.etal2016,Nishiura.etal2007} may help to cast our heuristic explanation of the localized eigenfunction, based on the concept of regional instability, into a more rigorous argument.
Similarly, such an approach may elucidate the transition from edge-localized peaks to boundary-localized peaks for too fast mass upregulation (cf.\ Appendix~\ref{app:non-adiabatic}, Fig.~\ref{fig:app_phase_diagram}).
In general, we expect that combining mathematical tools like singular perturbation theory with the local equilibria framework will be a fruitful approach to systematically study complex pattern-forming systems.

\begin{acknowledgments}

We thank Jacob Halatek and Korbinian P\"oppel  for discussions and feedback on the manuscript.
This research was supported by the Deutsche Forschungsgemeinschaft (DFG) via project A6 within the Collaborative Research Center ``Spatiotemporal dynamics of bacterial cells'' (TRR 174). 
We also gratefully acknowledge financial support by the DFG Research Training Group GRK2062 (Molecular Principles of Synthetic Biology). 
M.C.W. is supported by a DFG fellowship within the Graduate School of Quantitative Biosciences Munich.
\end{acknowledgments}

\appendix
\section{Reaction kinetics and template definition}
\label{app:model}

Throughout this paper we use a two-component McRD model on a one-dimensional domain with one protein species (cf.\ Eq.~\eqref{eq:system_eq1} and~\eqref{eq:system_eq2}). 
The proteins cycle between a cytosolic state (concentration $ c(x,t) $) and membrane bound state (concentration $ m(x,t) $) as specified by the reaction term $f(m,c)$.
Importantly, our results are based on the shape of the reactive nullclines and, hence, don't depend on the specific choice for $f(m,c)$. To illustrate our findings, we use biochemically motivated reaction kinetics, comprising attachment, $a(m)$, and detachment, $d(m)$, reactions
\begin{align}
 f(m,c) := a(m) c - d(m) m\ .
\end{align}
Specifically we use
\begin{subequations} \label{eq:model_specific}
 \begin{align}
  a(m)  &:= \left( k_\mathrm{on} + k_\mathrm{fb} \, m \right), \\
  d(m)  &:= \frac{k_{\mathrm{off}}}{K_{\mathrm{d}} + \, m},
 \end{align} 
\end{subequations}
as introduced before in Ref.~\cite{Brauns.etal2018}.
These reaction kinetics describe a protein species that can attach from the cytosol to the membrane with a rate $k_{\mathrm{on}}$ and get recruited from the cytosol to the membrane by membrane bound proteins with a rate $k_{\mathrm{fb}}$.
Membrane bound proteins detach from the membrane via an enzymatic process described by first order Michaelis-Menten kinetics, parameterized by the rate $k_{\mathrm{off}}$ and the dissociation constant $K_{\mathrm{d}}$.

We consider systems were the reaction rates are different in subdomains A and B.
This externally imposed heterogeneity was introduced as a step-like template in Sec.~\ref{sec:intro-model} (cf. Eq. \eqref{eq:template}).
For specificity, we choose a template that affects the reaction rates, such that the reactive nullcline in subdomain B is stretched along the $m$-axis with respect to the reactive nullcline in subdomain A.
To that end, we rescale the feedback rate and the dissociation constant scale with the template value, such that these rates become space dependent
\begin{subequations}
 \begin{align}
 	k_\mathrm{fb}(x)  & := \hat{k}_{\mathrm{fb}} / \temp(x),    \\
 	K_{\mathrm{d}}(x) & := \hat{K}_{\mathrm{d}} \, \temp(x).
 \end{align}
\end{subequations}%
The reaction term then becomes 
\begin{align}
\label{eq:m-scalnig}
 f(m,c; \theta) = 
  \left( k_\mathrm{on} + \hat{k}_{\mathrm{fb}} \, \frac{m}{\temp(x)}
  \right) \, c -  
  \frac{k_{\mathrm{off}} \, \frac{m}{\temp(x)}}{\hat{K}_\mathrm{d} + \frac{m}{\temp(x)}} 
 \;.
\end{align}

For convenience, we do not specify units of length and time. In an intracellular context a typical size would be $L \sim 10 \; \mu\text{m}$, and typical diffusion constants are $D_m \sim 0.01{-}0.1 \; \mu \text{m}^2 \text{s}^{-1}$ on the membrane and $D_c \sim 10 \; \mu \text{m}^2 \text{s}^{-1}$ in the cytosol. Rescaling to different spatial dimensions is straightforward. To fix a timescale, the kinetic rates can be rescaled with respect to the attachment rate $k_\mathrm{on}$. In an intracellular context, typical attachment timescales are on the order of seconds, i.e. $k_\mathrm{on} \sim \mathrm{s}^{-1}$.

\section{Nullclines without edge-sensing}
\label{app:diff_nullcline}

In Sec.~\ref{sec:regional-instability} in the main text, we analyzed the edge-sensing mechanism based on the reactive nullclines in phase space.
In this analysis, the effect of the heterogeneous reaction kinetics, i.e.\ the template, is captured by the shapes of the reactive nullclines. 
From our analysis, we found a criterion for the edge-sensing mechanism, as illustrated in \fig\ref{fig:regional_instability}.
The nullclines need to intersect at a point where only one nullcline is steeper than FBS (cf.\ Eq.~\eqref{eq:nc-slope-criterion}).

\begin{figure}
\includegraphics{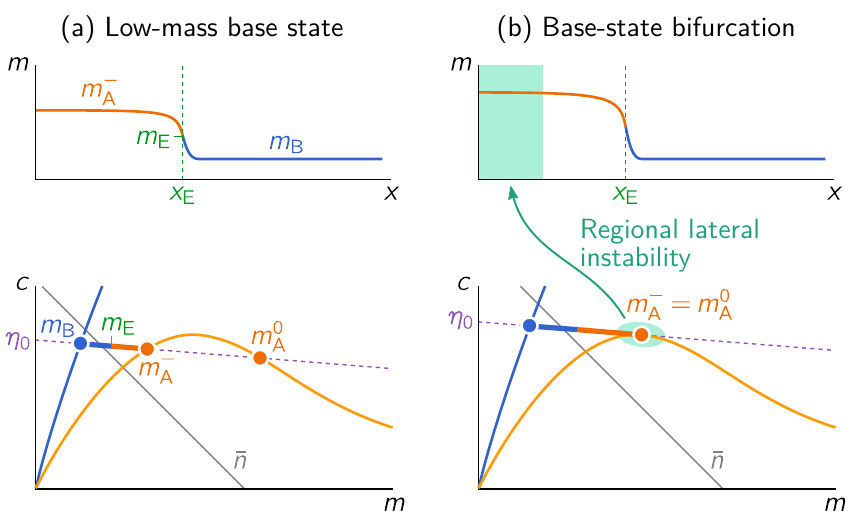}
\caption{\label{fig:app_role-of-nullcline-shapes} 
	Base states for nullclines shapes that do not facilitate edge-sensing. 
	(a) The low-mass base state with a high-concentration plateau in subdomain A and a low-concentration plateau in subdomain B.
	(b) At a critical average mass, the base state undergoes a saddle-node bifurcation. The bifurcation arises from the `annihilation' of the FBS-NC intersection point $m^-_A$ and $m^0_A$, and not from the break down of turnover balance as discussed in Sec.~\ref{sec:regional-instability} the main text. In this case, a regional instability is triggered at the system boundary.
	An adiabatic sweep of $\bar{n}$ through the saddle-node bifurcation of the FBS-NC intersection points is shown in Supplementary Movie~7.
}
\end{figure}

In this appendix, we discuss a case where the criterion for edge sensing is not fulfilled, such that a localization of the regional instability to the edge is not possible, and a peak at the template edge does not exist (cf.\ \fig\ref{fig:nullcline-criterion}(b)).
As an example, we consider a template, affecting the reaction kinetics such that the nullcline is stretched along the $c$-axes instead of the $m$-axis, as shown in \fig\ref{fig:app_role-of-nullcline-shapes}(b).

For the specific attachment--detachment reaction kinetics (Eq.~\eqref{eq:model_specific}), the nullclines is stretched along the $c$-axis via the off-rate while keeping all other rates constant:
\begin{equation}
 k_{\mathrm{off}}(x) := \hat{k}_{\mathrm{off}} \, \temp(x).
\end{equation}
The resulting reaction term then reads
\begin{align}
 f(m,c; \theta) = 
\left( k_\mathrm{on} + k_{\mathrm{fb}} \, m \right) \, c -  
  \theta(x) \frac{\hat{k}_{\mathrm{off}} \, m}{K_\mathrm{d} + m} .
\end{align}
Following the same arguments as presented in Sec.~\ref{sec:base_state},we can construct the base states (monotonic steady state).
Starting from the low-mass base state, illustrated in \fig\ref{fig:app_role-of-nullcline-shapes}(a), and increasing the average mass $ \bar{n} $ results in an upwards shift of the FBS by the same argument as presented in Sec.~\ref{sec:base_state}.
When the mass is further increased the FBS moves to the level where the two FBS-NC intersection points on the A-nullcline, $m_A^-$ and $m_A^0$, annihilate in a saddle-node bifurcation, as shown in \fig\ref{fig:app_role-of-nullcline-shapes}(b).
If $\eta_0$ increases beyond that point, the base state vanishes. 
Note that the origin of the saddle-node bifurcation lies in the annihilation of the two FBS-NC intersection points. 
This is different from the saddle-node bifurcation that occurs for the nullclines we analyzed in Sec.~\ref{sec:base_state} in the main text. 
There, the base state vanishes due to a breakdown of total turnover balance which becomes apparent by the `annihilation' of the two solutions for $\mE$ of Eq.~\eqref{eq:total-turnover}; cf.\ \fig\ref{fig:base_state}.

To study the dynamics in the vicinity of the base state bifurcation, we use the concept of regional instability (cf.\ Sec.~\ref{sec:regional-instability}).
The part of the density distribution that enters the laterally unstable region in phase space corresponds to the concentration at the left hand system boundary ($x=0$) (\fig\ref{fig:app_role-of-nullcline-shapes}(b)). 
Hence, upon crossing the base state's saddle-node bifurcation, a regional instability emerges at this system boundary, and a peak forms there.

Moreover, for nullclines shown in \fig\ref{fig:app_role-of-nullcline-shapes}, there is only one way to construct a stationary pattern with a single interface (inflection point) within subdomain A (in addition to the interface imposed by the template step).
This pattern always has a density peak at the system boundary ($ x = 0$) and decreases monotonically in space towards $x = L$. 
A stable peak localized to the template edge does not exist in this case.

\section{Temporal variation of the template}
\label{app:temp_template}
%
\begin{figure*}
 \includegraphics{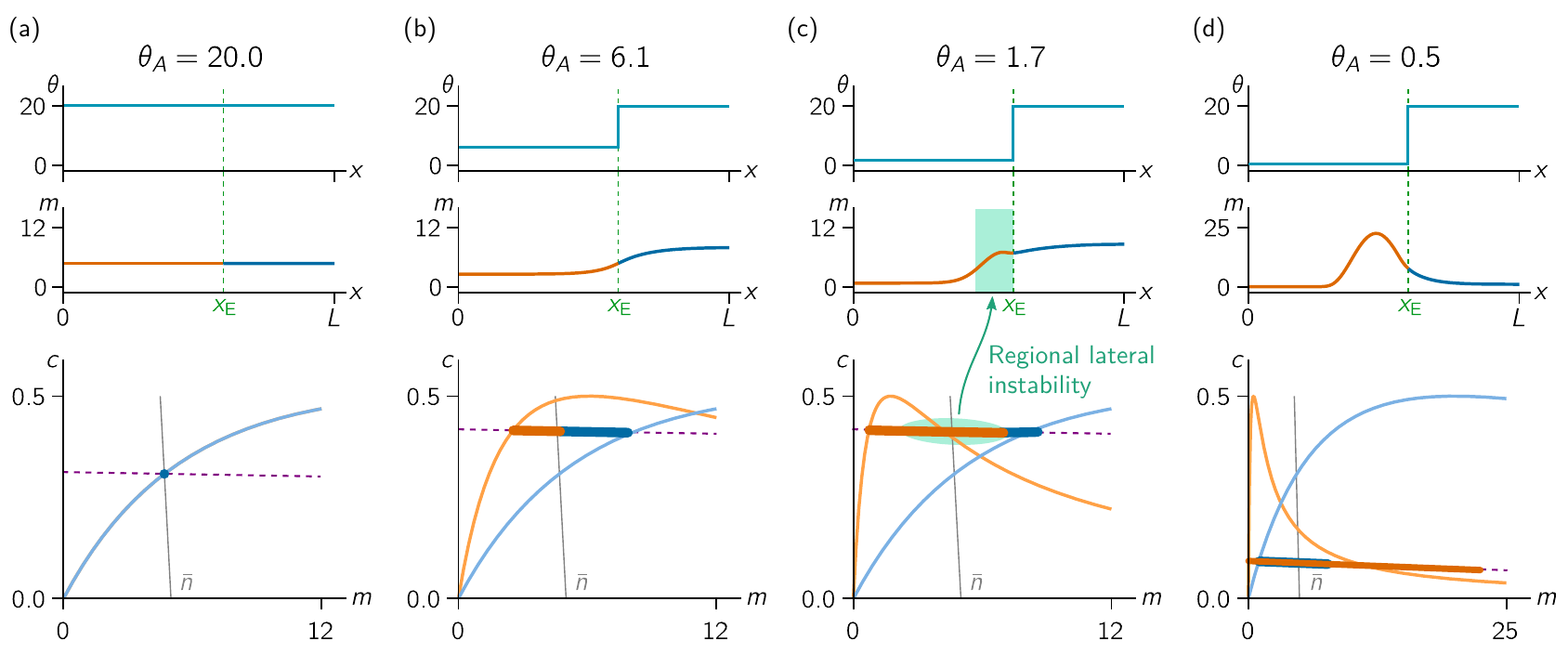}
 \caption{\label{fig:app_adiabatic_template} 
  An adiabatically changing template triggers peak formation at the template edge. See also Supplementary Movie~8.\\
    (a) Initial state with a homogeneous template profile $\temp_\mathrm{A} = \temp_\mathrm{B}$.
    (b) Base state analogous to \fig\ref{fig:base_state}. 
    (c) Base state right before the saddle-node bifurcation. 
        The region that becomes laterally unstable in the bifurcation is highlighted in light green in the spatial profile and in phase space (cf. \fig\ref{fig:regional_instability}).
  (d) Peak pattern state after the system transitioned through the bifurcation. This state is qualitatively the same as sketched in \fig\ref{fig:stationary-patterns}(c). \\
    Parameters as in \fig\ref{fig:moving_edge} but with $ L = 5 $, 
    $\theta_{\mathrm{B}} = 20 $, and
    $\theta_{\mathrm{A}}^{\mathrm{{f}}} = 0.5$.}
\end{figure*}

In Sec.~\ref{sec:steady-states} in the main text, we considered upregulation of average mass $\bar{n}$ while the spatial template was kept constant.
One can perform a similar analysis for a varying template, while keeping the average mass constant. 
In this scenario, not $ \bar{n} $ but the parametrization of the template (e.g.\ $\theta_\mathrm{A}$) serves as a bifurcation parameter.
In the biological context, this corresponds to a dynamically varying upstream protein pattern. 

To demonstrate that varying the template induces equivalent bifurcations as variation of the average mass, we use the template as in the main text (cf. Eq.~\eqref{eq:m-scalnig}) and perform a numeric simulation where we let the template $\theta(x,t)$ slowly vary with time (for simplicity only in subdomain A):
\begin{align}
 \temp(x,t) &= \begin{cases}
               \temp_{\mathrm{A}}(t) & x \leq \xE, \\
               \temp_{\mathrm{B}} & x > \xE.
              \end{cases} \; 
\end{align}
We initialize the template value in subdomain A at  $\theta_A(0) = \theta_B$ and let it change via a sinusoidal ramp to its final value  $\theta_A(t_\mathrm{f}) = \theta_A^\mathrm{f}$.

An exemplary simulation is shown in \fig\ref{fig:app_adiabatic_template} and Supplementary Movie 8.
At the start of the simulation, the reaction rates, and thus the reactive nullclines, are the same in subdomain A and subdomain B, which is illustrated by the overlapping nullclines in \fig\ref{fig:app_adiabatic_template}(a). The template is homogeneous, and the corresponding steady state is a homogeneous density profile for sufficiently low average mass.
Upon decreasing $\theta_\mathrm{A}$, the reaction rates in subdomain A change, leading to a change in the shape of the reactive nullcline (\fig\ref{fig:app_adiabatic_template}(b)). 
The resulting base state is equivalent to the low-mass base state, similar to the case for mass upregulation analyzed in Sec.~\ref{sec:base_state}. 
When $\theta_\mathrm{A}$ is further decreased, the density profile becomes regionally unstable at the template edge(\fig\ref{fig:app_adiabatic_template}(c)), which triggers a peak at the template edge (\fig\ref{fig:app_adiabatic_template}(d)).
This shows that the pattern formation process as discussed in the main text can also be realized with a dynamic template.

\section{Non-adiabatic mass upregulation}
\label{app:non-adiabatic}

In Sec.~\ref{sec:steady-states} in the main text, we analyzed the steady states as a function of average mass and found that the system undergoes a transition from base states to patterns through a saddle-node bifurcation. 
To analyze how the patterns emerge as the system goes through this bifurcation, we performed numerical simulations where we adiabatically increased the average mass by a global cytosolic source with rate $\kappa_\mathrm{s}$
\begin{equation} \label{eq:app_source}
	\partial_t c = D_c \, \partial_x^2 c - f(m,c) + \kappa_\mathrm{s},
\end{equation}
which entails $\partial_t \bar{n} = \kappa_\mathrm{s}$.
In this appendix, we explore the emergence of patterns when the average mass is non-adiabatically increased beyond the base-state bifurcation.
We initialize the system at zero mass and increase mass up to a value $\bar{n}_\mathrm{f}$ within the regime where no base state exists.
Varying the rate of mass inflow $\kappa_\mathrm{s}$,
we find six regimes with qualitatively different transient dynamics (see Supplementary Movies~9-14).
Below, we briefly describe the observed dynamics of the density profile in real space in these regimes going from slow to fast $\kappa_\mathrm{s}$.
We then use the $(m,c)$-phase space and the concept of regional instability (cf.\ Sec.~\ref{sec:regional-instability}) to heuristically explain the observed dynamics. 
 
 \begin{enumerate}
	\item \emph{Template-edge peak.}
For small $\kappa_\mathrm{s}$, a density peak emerges at the template edge, even though the density profile does not relax to a quasi-steady state. This highlights that the edge-sensing mechanism is robust against rate of mass inflow. (Supplementary Movie~9)
	\item \emph{Transition regime.}
Here, we observe two peaks emerging simultaneously, one at the outer boundary of subdomain A and one at the template edge.
This is an intermediate regime between the boundary peak regime (3) and the template edge peak regime  (1), as both peaks emerge simultaneously. (Supplementary Movie~10)
	\item \emph{System-boundary peak.}
Here, we observe one peak forming at the system boundary at $x=0$. (Supplementary Movie~11)
	\item \emph{Sequential peak formation.}
First, a peak forms at the domain boundary of subdomain A, as in the system- boundary peak regime (3).
Then, after the first peak already formed, another peak forms at the template edge as in regime (1).(Supplementary Movie~12)
	\item \emph{Multiple peaks.}
Here, the pattern-formation process is very similar to the system-boundary peak regime (3).
However, multiple peaks form at the outer boundary of subdomain A simultaneously. (Supplementary Movie~13)
	\item \emph{Quenched system.}
Here the mass is upregulated almost instantaneously, $\kappa_\mathrm{s}^{-1} {\rightarrow} \, 0$, which is equivalent to a system initialized with the complete mass in the cytosol. This results in a sequence of peaks forms in subdomain A, starting from the template edge in a process akin to front invasion into an unstable state \cite{vanSaarloos2003}. (Supplementary Movie~14)
\end{enumerate}
Note that all states with multiple peaks (corresponding to multiple inflection points of the pattern profile in subdomain A) are unstable due to coarsening. The final steady state is always a pattern with a peak either at the template edge or at the system boundary $x = 0$.

\begin{figure}
 \includegraphics{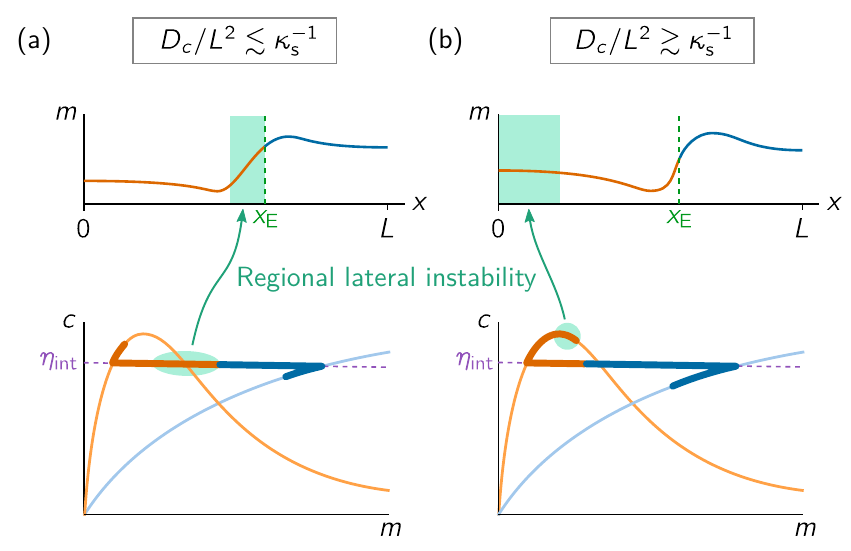}
 \caption{\label{fig:app_non-adiabatic}
 Effect of non-adiabatic mass upregulation on the pattern formation dynamics.
  (a) Peak formation at the template edge: The density distribution is not embedded in a single FBS, leading to the `zig-zag'-shaped density distribution in phase space. The regional instability is still triggered at the template edge, as highlighted by the  (green) shaded region.
  (b) Peak formation at the system boundary: The faster mass-inflow leads to a more pronounced `zig-zag'-shaped density distribution in phase space. The regional instability is now first triggered at the system boundary.
  This results in a peak forming at the system boundary as shown in \fig\ref{fig:stationary-patterns}(b).}
\end{figure}

In Sec.~\ref{sec:regional-instability} and Appendix \ref{app:diff_nullcline}, we showed that the formation of a density peak is determined by the position of a regional instability, when the system is in quasi-steady state.
The position of the regional instability can be found from a phase portrait analysis, since the nullcline slope criterion (cf.\ Eq.~\eqref{eq:nc-slope-criterion}) determines which part of the density profile becomes unstable.
To use the same heuristic for understanding the formation of these transient peaks, we analyze the density distribution in phase space obtained from numerical simulations in these non-adiabatic regimes.
 When mass is added to the system on a faster timescale than it can relax to its steady state, the density distribution in phase space is no longer embedded in a single FBS. 
 Instead, the density distribution in phase space follows a `zig-zag' shape, as illustrated in \fig\ref{fig:app_non-adiabatic}.
 This indicates that the density in the vicinity of the template edge is still embedded in a FBS, with offset $\eta_\mathrm{int}$, but the density far away from the template edge deviates from this FBS. 
Instead, these `quasi-plateaus' are slaved to the nullcline which indicates that they are locally close to reactive equilibrium, and that their relaxation is limited by diffusive mass transport.
Accordingly, for faster inflow of mass, the zig-zag shape is more pronounced---that is, the quasi-plateaus deviate more from $\eta_\mathrm{int}$.
If inflow of mass is faster than diffusive transport across the quasi-plateaus, a region at the system boundary in subdomain A enters the lateral unstable region in phase space first, as illustrated in \fig\ref{fig:app_non-adiabatic}(b).
This leads to the emergence of a peak at the system boundary in regime (3), and to more complex pattern formation in regimes (4)--(6). In the transition regime (2), mass inflow and mass transport roughly balance, such that a region at the system boundary and a region at the template edge enter the laterally unstable region in phase space at the same time.
 
\begin{figure}
 \includegraphics{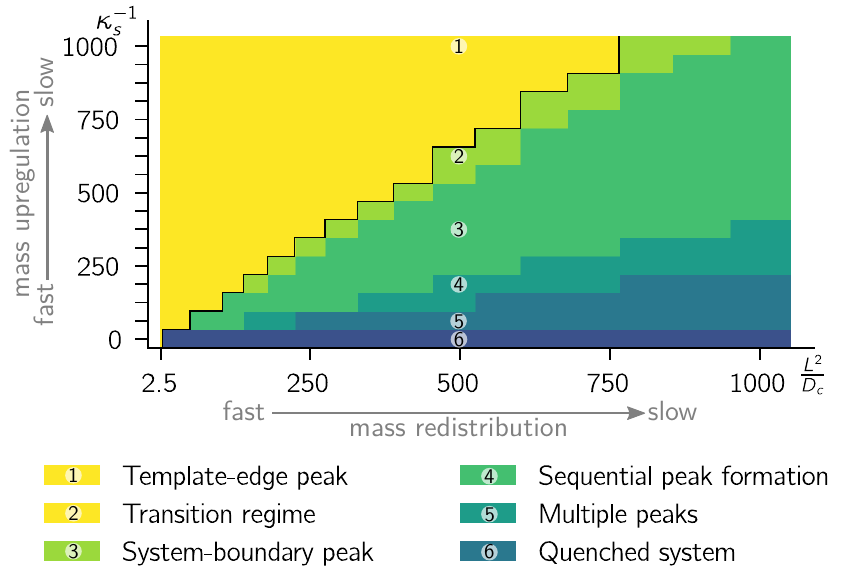}%
 \caption{\label{fig:app_phase_diagram}
Phase diagram for the timescale of mass upregulation (global cytosolic inflow $\kappa_\mathrm{s}^{-1}$) against the timescale of mass-redistribution across the entire domain ($L^2/D_c$).
Edge sensing, i.e.\ formation of a single density peak a the template edge, is operational in regime (1), see Supplementary Movie~9. In regimes (2)-(6) a peak at the system boundary ($x = 0$) or multiple peaks form, see Supplementary Movies 10-14. \\
Parameters:
  $D_m = 0.01$, 
  $D_c = 10$, 
  $k_{\mathrm{on}}       = 1$, 
  $\hat{k}_{\mathrm{fb}} = 1$, 
  $k_{\mathrm{off}}      = 2$, 
  $\hat{K}_{\mathrm{d}}  = 1$, 
  $\theta_{\mathrm{A}} = 0.5$, 
  $\theta_{\mathrm{B}} = 20$, 
  $x_E  = 3/5 \, L$,
  $\bar{n}_{\mathrm{f}} = 5$.
}
\end{figure}

The $(L^2/D_c, \kappa_\mathrm{s}^{-1})$-phase diagram shown in \fig\ref{fig:app_phase_diagram} confirms the intuition that pattern emergence depends on the competition between the time scales of mass inflow, $\kappa_\mathrm{s}$, and diffusive mass transport across the system ${\sim}L^2/D_c$.
Indeed, the regime boundaries in the phase diagram are roughly straight lines emanating from the origin. 
In particular, the transition from a `template-edge peak' to a `system-boundary peak' corresponds to a line $\kappa_\mathrm{s}^{-1} \approx L^2/D_c$.
This confirms the intuition that edge sensing is only possible when the inflow of mass into the system is slower than the timescale of diffusive mass transport. 
For comparison, in an intracellular context one has $L \approx 10 \; \mu \mathrm{m}$ and $D_c \approx 10 \; \mu \mathrm{m}^2 \mathrm{s}^{-1}$, such that the  timescale of mass transport across the cell is on the order of seconds.
This is fast compared to changes in average protein concentrations (for instance, due to protein expression or release from the nucleus). 
Hence, the edge-sense mechanism is a realistic candidate for template guided intracellular pattern formation. 

\section{Numerical continuation, steady state construction and bifurcation scenarios}
\label{app:continuation}
\textit{Numerical continuation.}\;---\;%
To numerically calculate steady state solutions of the two-component McRD system (Eq.~\eqref{eq:rd-system}), we choose a finite-difference discretization of the PDEs.
For steady states, this yields an algebraic system of equations that can be solved with an iterative Newton method.
The basic idea of numerical continuation is to follow a solution branch through parameter space (see for instance, Ref.~\cite{Krauskopf.etal2007} for an excellent overview over continuation methods). This ``path-following'' is often performed by emplying a predictor-–corrector scheme: 
Starting from one solution, the next solution along the branch is predicted from the tangent space of the solution branch which can be obtained from the Jacobian. 

\begin{figure}
 \includegraphics{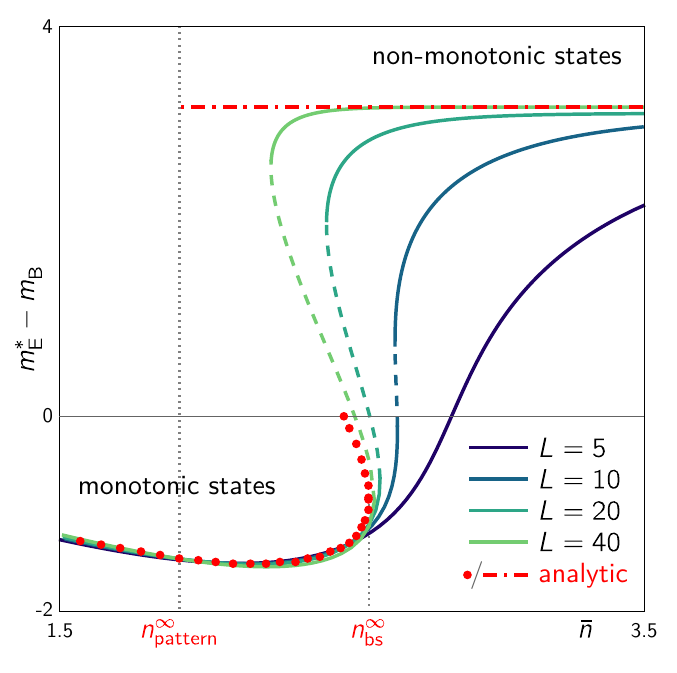}
 \caption{\label{fig:base-state-systemsize}
One-parameter bifurcation structure in $\bar{n}$ (average mass) connecting the low-mass base state (monotonic, i.e.\ $\mE < \mB$ and the peak pattern at the template edge (non-monotonic, i.e.\ $\mE > \mB$). Solid (dashed) lines indicate stable (unstable) branches from numerical continuation for different domain sizes (increasing from dark to light lines). Solutions from the analytic construction of base states in the large domain size limit ($L \rightarrow \infty$) are shown as red dots and  (position of the saddle-node bifurcation denoted by $n_\mathrm{bs}^\infty$). Note that for small domain size ($L = 5$), the saddle-node bifurcations vanish and the base state smoothly transitions into a stable peak pattern upon increasing $\bar{n}$.
The red, dash dotted line indicates the analytically constructed edge-localized pattern (limit $L \rightarrow \infty$). The lower bound in average mass for the existence of these patterns is denoted by $n_\mathrm{pattern}^\infty$. \\
Fixed parameters as in \fig\ref{fig:stationary-patterns}(a).
}
\end{figure}

\textit{Steady state construction and finite domain size effects.}\;---\;%
In order to test the geometric constructions introduced in Sec.~\ref{sec:steady-states} we characterize the steady state of the system with the quantity $\mEeq - \mB$, which must be negative for low-mass base states (monotonic steady states in the low-mass regime, cf. Sec.~\ref{sec:base_state}) and positive for non-monotonic steady states (stationary peak pattern localized at the template edge, cf. Sec.~\ref{sec:stationary}). We perform numerical continuation and compare the results from the simulation (solid lines in \fig\ref{fig:base-state-systemsize}) with the approximation from the geometric construction (red dots and dash-dotted line). The geometric construction serves as a good approximation for the steady for sufficiently large system sizes.

Also note that for small system sizes the base state smoothly transitions into the pattern state (purple line corresponding to $L = 5$ in \fig\ref{fig:base-state-systemsize}).

\textit{Bifurcation scenario.}\;---\;%
The bifurcation scenario connecting the base state and the patterns can be understood as a series of imperfect subcritical pitchfork bifurcations. The imperfection is caused by the template that breaks mirror symmetry of the system. On a homogeneous domain (i.e.\ without a template), the bifurcations from homogeneous steady state to patterns are subcritical pitchfork bifurcation that become supercritical for small system sizes / large wavenumbers \cite{Brauns.etal2018}.
A more detailed analysis of the bifurcation scenario induced by the step-like template is left for future work.
One interesting starting point would be to analyze the two-parameter bifurcation diagram in the $
(\bar{n},\theta_\mathrm{A})$-plane, where the line $\theta_\mathrm{A} = \theta_\mathrm{B}$ correspond to the homogeneous domain. 
Alternatively, one can investigate the bifurcation scenario in the template edge position (i.e.\ the $(\bar{n},\xE)$ parameter plane), where $\xE = 0$ and $\xE = L$ correspond to homogeneous domains. 

\begin{figure}
\includegraphics{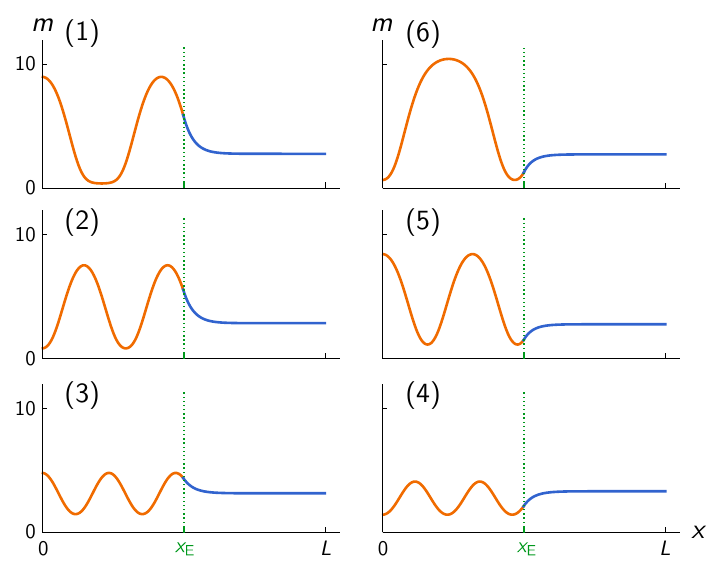}
\caption{\label{fig:unstable-states} 
	Spatial profiles representative of the unstable branches numbered (1-6) in the $\bar{n}$-bifurcation diagram Fig.~\ref{fig:stationary-patterns}(a). 
}
\end{figure}

\textit{Unstable multi-interface patterns.}\;---\;%
The dotted branches in the bifurcation structure \fig\ref{fig:stationary-patterns}(a) correspond to patterns with multiple self-organized interfaces (i.e.\ more than two inflection points in the spatial profile, since the template edge enforces one inflection point at $\xE$).
Figure~\ref{fig:unstable-states} shows spatial profiles at the numbered points in \fig\ref{fig:stationary-patterns}(a) representative for the respective branches.
The spiral structure of the bifurcation structure reflects an increasing number of of peaks from the outside to the center of the spiral.
For the branches numbered 1-3 (4-6) the concentration difference $\mEeq - \mB$ is positive (negative), corresponding to a peak (trough) at the template edge. 

All multi-interface patterns are unstable due to a competition for mass and decay into one of the two stable patterns, with a peak either at the system boundary or at the template edge, in a coarsening process. 

\section{Demonstration of the edge-sensing criterion}
\label{app:cdc42}
\def\kon{k_\mathrm{on}}
\def\kfb{k_\mathrm{fb}}
\def\koff{k_\mathrm{off}}
\def\gon{\gamma_\mathrm{on}}
\def\goff{\gamma_\mathrm{off}}

\begin{figure*}
 \includegraphics{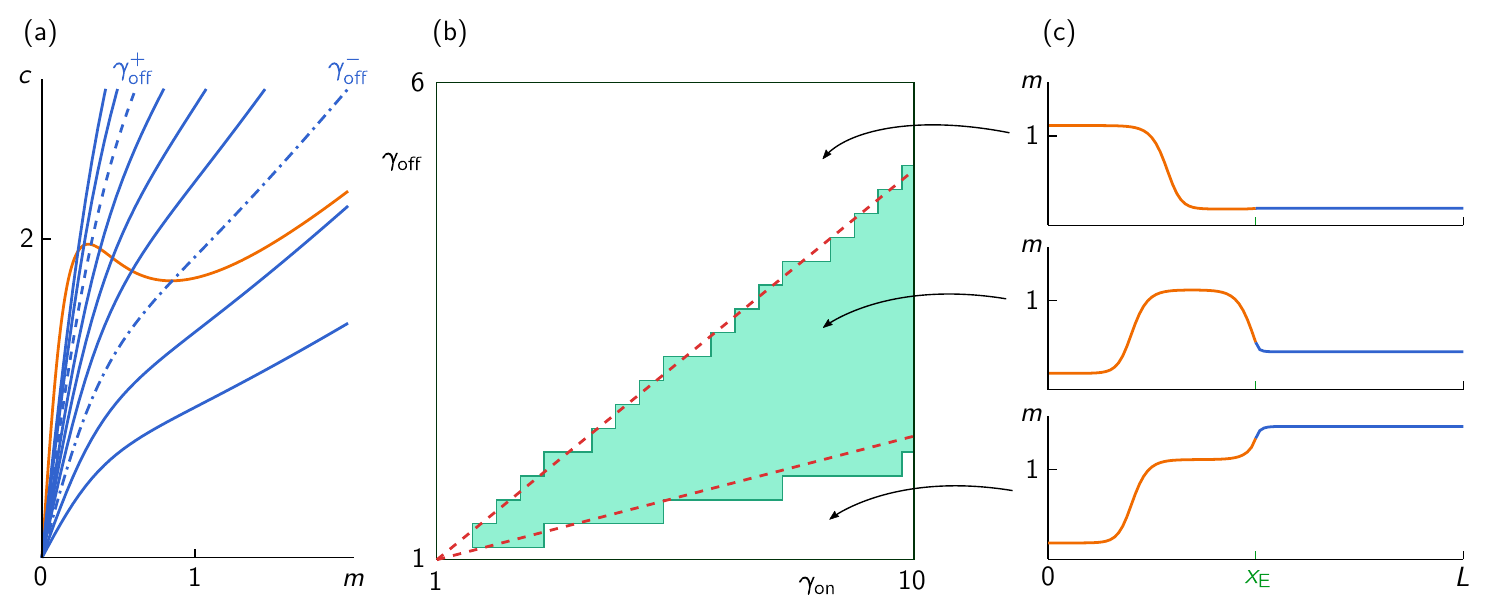}
 \caption{\label{fig:Cdc42-model-example}
Demonstration of the edge-sensing criterion for a phenomenological model for Cdc42 pattern formation.
(a) Nullclines of the reaction kinetics Eq.~\eqref{eq:Cdc42-model} in subdomains~A (orange) and~B (blue), defined by the reaction rates in Eq.~\eqref{eq:Cdc42-AB-rates}. Multiple B-nullclines are shown for off-rate factors $\goff = 1,1.5, \ldots, 5$ while the on-rate factor is fixed at $\gon = 8$. The critical values $\goff^\pm$ between which the A-nullcline is intersected at negative slope (i.e.\ the edge-sensing criterion is fulfilled) are shown as dashed and dash-dotted lines respectively.
(b) Phase diagram of on- and off-rate factors $(\gon,\goff)$. The shaded region shows the regime where edge-sensing is observed in numerical simulations with adiabatically increasing average mass (cf.\ Eq.~\eqref{eq:app_source}). Dashed red lines show the curves $\goff^\pm(\gon)$, cf.\ Eq.~\eqref{eq:Cdc42-goff-critical}, between which the edge-sensing criterion is fulfilled.
(c) Quasi-stationary density profiles obtained from numerical simulations with adiabatically increasing average mass $\bar{n}$ illustrating the typical pattern emerging from the base-states's saddle-node bifurcation in the three regimes of the $(\gon,\goff)$ phase diagram ($\gon = 8$; top: $\goff = 5.5, \bar{n} = 2.9$; $\goff = 3, \bar{n} = 2.35$; bottom: $\goff = 1.5, \bar{n} = 2.24$.). 
Fixed parameters: $\kfb = 1, \kon = 0.07, \koff = 1, D_m = 10^{-4}, D_c = 0.1, L = 1, \xE = 0.5, \kappa_\mathrm{s} = 10^{-5}$.
}
\end{figure*}

To illustrate how the edge sensing criterion might help in modeling of biological systems we consider a concrete biological example. 
Cdc42 pattern formation is described by a phenomenological two-component model of the form Eq.~\eqref{eq:rd-system} with the reaction kinetics \cite{Mori.etal2008}
\begin{equation} \label{eq:Cdc42-model}
	f(m,c) = \left(\kon + \kfb \frac{m^2}{1+m^2}\right) c - \koff m
\end{equation}

The nullcline $c^*(m)$ determined by $f(m, c^*(m)) = 0$ has a section of negative slope for $\kfb > 8\kon$ in the interval $[m_\mathrm{max}, m_\mathrm{min}]$ given by
\begin{equation} \label{app-eq:m-minmax}
	m_\mathrm{min,max} = \sqrt{\frac{1}{1+\kon/\kfb} \left( \frac{1 \mp \nu}{2} - \frac{\kon}{\kfb} \right)}
\end{equation}
with
\begin{equation}
	\nu = \sqrt{1 - 8 \frac{\kon}{\kfb}}.	
\end{equation}

Since cytosolic diffusion is multiple orders of magnitude faster than diffusion of membrane bound Cdc42, we consider the limit $D_c \gg D_m$ where the FBS-slope is zero. Then the criterium for lateral instability is that the nullcline slope be negative $\partial_m c^*(m) < 0$; and the edge-sensing criterion is that nullclines must intersect at a point where one of them has negative slope.

In cellular wound healing, Abr forms a ring of high concentration around the wound edge, followed by the formation of a Cdc42 ring around the Abr ring \cite{Vaughan.etal2011}. Cdc42 then activates the actomyosin machinery that drives the contraction of the cell membrane to close the wound.
Protein mutation on Abr and Cdc42 studies suggest that the high density Abr ring acts as a template for Cdc42 \cite{Vaughan.etal2011}.
Furthermore, it has been shown that Abr acts as both a nucleotide exchange factor (GEF) and a GTPase-activating protein (GAP) for Cdc42 \cite{Vaughan.etal2011}. 
We therefore model the effect of Abr on the Cdc42 kinetics as a factor increasing both the attachment rate $\kon$ and the detachment rate $\koff$.
The reaction terms in the two subdomains with low and high Abr density, are thus defined as $f_\mathrm{A,B}(m,c) = f(m,c; \kon^\mathrm{A,B} ,\koff^\mathrm{A,B})$ with 
\begin{align} \label{eq:Cdc42-AB-rates}
	\kon^\mathrm{A} &= \kon, &\; \kon^\mathrm{B} &= \gon \kon, \\
	\koff^\mathrm{A} &= \koff, &\; \koff^\mathrm{B} &= \goff \koff,
\end{align}
where the factors $\gon, \goff > 1$ encode the relative enhancement of attachment and detachment rates in subdomain~B compared to subdomain~A.

The edge-sensing criterion (nullclines intersect at a point where the A-nullcline has negative slope) is fulfilled when the intersection point $(m_\mathrm{i},c_\mathrm{i})$ of the two nullclines, defined by $\fA(m_\mathrm{i},c_\mathrm{i}) = 0 \; \& \; \fB(m_\mathrm{i},c_\mathrm{i}) = 0$, lies in the range $[m_\mathrm{max}, m_\mathrm{min}]$ (see Fig.~\ref{fig:Cdc42-model-example}(a)).
Solving these equations for $\goff$, we find that nullclines fulfill the edge-sensing criterion when $\goff$ lies in the range $[\goff^-,\goff^+]$ as a function of $\gon$:
\begin{equation} \label{eq:Cdc42-goff-critical}
	\goff^\pm(\gon) = \frac{1}{1+ \kfb/\kon} \left( \frac{\kfb}{\kon} + \frac{2}{\nu \mp 1} + \gon \frac{\nu \mp 3}{\nu \mp 1} \right).
\end{equation}
Note that $\goff^\pm(1) = 1$ corresponds to the singular case where the template has no effect, i.e.\ the reaction kinetics in the two subdomains are identical. 
The dashed, red lines in Fig.~\ref{fig:Cdc42-model-example}(b) show the curves $\goff^\pm(\gon)$ that delineate the regime where the edge-sensing criterion is fulfilled.

To test the criterion, we ran numerical simulations, for different combinations $(\gon,\goff)$. In each simulation, the average mass is adiabatically increased at a rate $\kappa_\mathrm{s} = 10^{-5}$, cf.\ Eq.~\eqref{eq:app_source}.

We find three different types of patterns emerging from the base-states's saddle-node bifurcation. Examples of these are shown in Fig.~\ref{fig:Cdc42-model-example}(c). The regime where we find edge-localized peaks agrees well with the prediction from the geometric edge-sensing criterion (shown as dashed, red lines in (b)).

Interestingly, the $(\gon,\goff)$-phase diagram shows that both attachment and detachment need to be enhanced in subdomain~B. This implies that proteins, like Abr, that have both GEF- and GAP-catalytic domains may play a crucial role for edge sensing by GTPases.

\newpage


\begin{thebibliography}{53}%
\makeatletter
\providecommand \@ifxundefined [1]{%
 \@ifx{#1\undefined}
}%
\providecommand \@ifnum [1]{%
 \ifnum #1\expandafter \@firstoftwo
 \else \expandafter \@secondoftwo
 \fi
}%
\providecommand \@ifx [1]{%
 \ifx #1\expandafter \@firstoftwo
 \else \expandafter \@secondoftwo
 \fi
}%
\providecommand \natexlab [1]{#1}%
\providecommand \enquote  [1]{``#1''}%
\providecommand \bibnamefont  [1]{#1}%
\providecommand \bibfnamefont [1]{#1}%
\providecommand \citenamefont [1]{#1}%
\providecommand \href@noop [0]{\@secondoftwo}%
\providecommand \href [0]{\begingroup \@sanitize@url \@href}%
\providecommand \@href[1]{\@@startlink{#1}\@@href}%
\providecommand \@@href[1]{\endgroup#1\@@endlink}%
\providecommand \@sanitize@url [0]{\catcode `\\12\catcode `\$12\catcode
  `\&12\catcode `\#12\catcode `\^12\catcode `\_12\catcode `\%12\relax}%
\providecommand \@@startlink[1]{}%
\providecommand \@@endlink[0]{}%
\providecommand \url  [0]{\begingroup\@sanitize@url \@url }%
\providecommand \@url [1]{\endgroup\@href {#1}{\urlprefix }}%
\providecommand \urlprefix  [0]{URL }%
\providecommand \Eprint [0]{\href }%
\providecommand \doibase [0]{http://dx.doi.org/}%
\providecommand \selectlanguage [0]{\@gobble}%
\providecommand \bibinfo  [0]{\@secondoftwo}%
\providecommand \bibfield  [0]{\@secondoftwo}%
\providecommand \translation [1]{[#1]}%
\providecommand \BibitemOpen [0]{}%
\providecommand \bibitemStop [0]{}%
\providecommand \bibitemNoStop [0]{.\EOS\space}%
\providecommand \EOS [0]{\spacefactor3000\relax}%
\providecommand \BibitemShut  [1]{\csname bibitem#1\endcsname}%
\let\auto@bib@innerbib\@empty
\bibitem [{\citenamefont {Halatek}\ \emph {et~al.}(2018)\citenamefont
  {Halatek}, \citenamefont {Brauns},\ and\ \citenamefont
  {Frey}}]{Halatek.etal2018}%
  \BibitemOpen
  \bibfield  {author} {\bibinfo {author} {\bibfnamefont {J.}~\bibnamefont
  {Halatek}}, \bibinfo {author} {\bibfnamefont {F.}~\bibnamefont {Brauns}}, \
  and\ \bibinfo {author} {\bibfnamefont {E.}~\bibnamefont {Frey}},\ }\href
  {\doibase 10.1098/rstb.2017.0107} {\bibfield  {journal} {\bibinfo  {journal}
  {Philosophical Transactions of the Royal Society B: Biological Sciences}\
  }\textbf {\bibinfo {volume} {373}},\ \bibinfo {pages} {20170107} (\bibinfo
  {year} {2018})}\BibitemShut {NoStop}%
\bibitem [{\citenamefont {Howard}\ \emph {et~al.}(2001)\citenamefont {Howard},
  \citenamefont {Rutenberg},\ and\ \citenamefont {{de Vet}}}]{Howard.etal2001}%
  \BibitemOpen
  \bibfield  {author} {\bibinfo {author} {\bibfnamefont {M.}~\bibnamefont
  {Howard}}, \bibinfo {author} {\bibfnamefont {A.~D.}\ \bibnamefont
  {Rutenberg}}, \ and\ \bibinfo {author} {\bibfnamefont {S.}~\bibnamefont {{de
  Vet}}},\ }\href {\doibase 10.1103/PhysRevLett.87.278102} {\bibfield
  {journal} {\bibinfo  {journal} {Physical Review Letters}\ }\textbf {\bibinfo
  {volume} {87}} (\bibinfo {year} {2001}),\
  10.1103/PhysRevLett.87.278102}\BibitemShut {NoStop}%
\bibitem [{\citenamefont {Huang}\ \emph {et~al.}(2003)\citenamefont {Huang},
  \citenamefont {Meir},\ and\ \citenamefont {Wingreen}}]{Huang.etal2003}%
  \BibitemOpen
  \bibfield  {author} {\bibinfo {author} {\bibfnamefont {K.~C.}\ \bibnamefont
  {Huang}}, \bibinfo {author} {\bibfnamefont {Y.}~\bibnamefont {Meir}}, \ and\
  \bibinfo {author} {\bibfnamefont {N.~S.}\ \bibnamefont {Wingreen}},\ }\href
  {\doibase 10.1073/pnas.2135445100} {\bibfield  {journal} {\bibinfo  {journal}
  {Proceedings of the National Academy of Sciences}\ }\textbf {\bibinfo
  {volume} {100}},\ \bibinfo {pages} {12724} (\bibinfo {year}
  {2003})}\BibitemShut {NoStop}%
\bibitem [{\citenamefont {Mori}\ \emph {et~al.}(2008)\citenamefont {Mori},
  \citenamefont {Jilkine},\ and\ \citenamefont
  {{Edelstein-Keshet}}}]{Mori.etal2008}%
  \BibitemOpen
  \bibfield  {author} {\bibinfo {author} {\bibfnamefont {Y.}~\bibnamefont
  {Mori}}, \bibinfo {author} {\bibfnamefont {A.}~\bibnamefont {Jilkine}}, \
  and\ \bibinfo {author} {\bibfnamefont {L.}~\bibnamefont
  {{Edelstein-Keshet}}},\ }\href {\doibase 10.1529/biophysj.107.120824}
  {\bibfield  {journal} {\bibinfo  {journal} {Biophysical Journal}\ }\textbf
  {\bibinfo {volume} {94}},\ \bibinfo {pages} {3684} (\bibinfo {year}
  {2008})}\BibitemShut {NoStop}%
\bibitem [{\citenamefont {Goryachev}\ and\ \citenamefont
  {Pokhilko}(2008)}]{Goryachev.Pokhilko2008}%
  \BibitemOpen
  \bibfield  {author} {\bibinfo {author} {\bibfnamefont {A.~B.}\ \bibnamefont
  {Goryachev}}\ and\ \bibinfo {author} {\bibfnamefont {A.~V.}\ \bibnamefont
  {Pokhilko}},\ }\href {\doibase 10.1016/j.febslet.2008.03.029} {\bibfield
  {journal} {\bibinfo  {journal} {FEBS Letters}\ }\textbf {\bibinfo {volume}
  {582}},\ \bibinfo {pages} {1437} (\bibinfo {year} {2008})}\BibitemShut
  {NoStop}%
\bibitem [{\citenamefont {Ishihara}\ \emph {et~al.}(2007)\citenamefont
  {Ishihara}, \citenamefont {Otsuji},\ and\ \citenamefont
  {Mochizuki}}]{Ishihara.etal2007}%
  \BibitemOpen
  \bibfield  {author} {\bibinfo {author} {\bibfnamefont {S.}~\bibnamefont
  {Ishihara}}, \bibinfo {author} {\bibfnamefont {M.}~\bibnamefont {Otsuji}}, \
  and\ \bibinfo {author} {\bibfnamefont {A.}~\bibnamefont {Mochizuki}},\ }\href
  {\doibase 10.1103/PhysRevE.75.015203} {\bibfield  {journal} {\bibinfo
  {journal} {Physical Review E}\ }\textbf {\bibinfo {volume} {75}} (\bibinfo
  {year} {2007}),\ 10.1103/PhysRevE.75.015203}\BibitemShut {NoStop}%
\bibitem [{\citenamefont {Otsuji}\ \emph {et~al.}(2007)\citenamefont {Otsuji},
  \citenamefont {Ishihara}, \citenamefont {Co}, \citenamefont {Kaibuchi},
  \citenamefont {Mochizuki},\ and\ \citenamefont {Kuroda}}]{Otsuji.etal2007}%
  \BibitemOpen
  \bibfield  {author} {\bibinfo {author} {\bibfnamefont {M.}~\bibnamefont
  {Otsuji}}, \bibinfo {author} {\bibfnamefont {S.}~\bibnamefont {Ishihara}},
  \bibinfo {author} {\bibfnamefont {C.}~\bibnamefont {Co}}, \bibinfo {author}
  {\bibfnamefont {K.}~\bibnamefont {Kaibuchi}}, \bibinfo {author}
  {\bibfnamefont {A.}~\bibnamefont {Mochizuki}}, \ and\ \bibinfo {author}
  {\bibfnamefont {S.}~\bibnamefont {Kuroda}},\ }\href {\doibase
  10.1371/journal.pcbi.0030108} {\bibfield  {journal} {\bibinfo  {journal}
  {PLoS Computational Biology}\ }\textbf {\bibinfo {volume} {3}},\ \bibinfo
  {pages} {e108} (\bibinfo {year} {2007})}\BibitemShut {NoStop}%
\bibitem [{\citenamefont {Halatek}\ and\ \citenamefont
  {Frey}(2012)}]{Halatek.Frey2012}%
  \BibitemOpen
  \bibfield  {author} {\bibinfo {author} {\bibfnamefont {J.}~\bibnamefont
  {Halatek}}\ and\ \bibinfo {author} {\bibfnamefont {E.}~\bibnamefont {Frey}},\
  }\href {\doibase 10.1016/j.celrep.2012.04.005} {\bibfield  {journal}
  {\bibinfo  {journal} {Cell Reports}\ }\textbf {\bibinfo {volume} {1}},\
  \bibinfo {pages} {741} (\bibinfo {year} {2012})}\BibitemShut {NoStop}%
\bibitem [{\citenamefont {Kl{\"u}nder}\ \emph {et~al.}(2013)\citenamefont
  {Kl{\"u}nder}, \citenamefont {Freisinger}, \citenamefont
  {{Wedlich-S{\"o}ldner}},\ and\ \citenamefont {Frey}}]{Klunder.etal2013}%
  \BibitemOpen
  \bibfield  {author} {\bibinfo {author} {\bibfnamefont {B.}~\bibnamefont
  {Kl{\"u}nder}}, \bibinfo {author} {\bibfnamefont {T.}~\bibnamefont
  {Freisinger}}, \bibinfo {author} {\bibfnamefont {R.}~\bibnamefont
  {{Wedlich-S{\"o}ldner}}}, \ and\ \bibinfo {author} {\bibfnamefont
  {E.}~\bibnamefont {Frey}},\ }\href {\doibase 10.1371/journal.pcbi.1003396}
  {\bibfield  {journal} {\bibinfo  {journal} {PLOS Computational Biology}\
  }\textbf {\bibinfo {volume} {9}},\ \bibinfo {pages} {e1003396} (\bibinfo
  {year} {2013})}\BibitemShut {NoStop}%
\bibitem [{\citenamefont {Trong}\ \emph {et~al.}(2014)\citenamefont {Trong},
  \citenamefont {Nicola}, \citenamefont {Goehring}, \citenamefont {Kumar},\
  and\ \citenamefont {Grill}}]{Trong.etal2014}%
  \BibitemOpen
  \bibfield  {author} {\bibinfo {author} {\bibfnamefont {P.~K.}\ \bibnamefont
  {Trong}}, \bibinfo {author} {\bibfnamefont {E.~M.}\ \bibnamefont {Nicola}},
  \bibinfo {author} {\bibfnamefont {N.~W.}\ \bibnamefont {Goehring}}, \bibinfo
  {author} {\bibfnamefont {K.~V.}\ \bibnamefont {Kumar}}, \ and\ \bibinfo
  {author} {\bibfnamefont {S.~W.}\ \bibnamefont {Grill}},\ }\href {\doibase
  10.1088/1367-2630/16/6/065009} {\bibfield  {journal} {\bibinfo  {journal}
  {New Journal of Physics}\ }\textbf {\bibinfo {volume} {16}},\ \bibinfo
  {pages} {065009} (\bibinfo {year} {2014})}\BibitemShut {NoStop}%
\bibitem [{\citenamefont {Alonso}\ and\ \citenamefont
  {B{\"a}r}(2014)}]{Alonso.Bar2014}%
  \BibitemOpen
  \bibfield  {author} {\bibinfo {author} {\bibfnamefont {S.}~\bibnamefont
  {Alonso}}\ and\ \bibinfo {author} {\bibfnamefont {M.}~\bibnamefont
  {B{\"a}r}},\ }\href {\doibase 10.1140/epjnbp14} {\bibfield  {journal}
  {\bibinfo  {journal} {EPJ Nonlinear Biomed Phys}\ }\textbf {\bibinfo {volume}
  {2}},\ \bibinfo {pages} {1} (\bibinfo {year} {2014})}\BibitemShut {NoStop}%
\bibitem [{\citenamefont {Wu}\ \emph {et~al.}(2016)\citenamefont {Wu},
  \citenamefont {Halatek}, \citenamefont {Reiter}, \citenamefont {Kingma},
  \citenamefont {Frey},\ and\ \citenamefont {Dekker}}]{Wu.etal2016}%
  \BibitemOpen
  \bibfield  {author} {\bibinfo {author} {\bibfnamefont {F.}~\bibnamefont
  {Wu}}, \bibinfo {author} {\bibfnamefont {J.}~\bibnamefont {Halatek}},
  \bibinfo {author} {\bibfnamefont {M.}~\bibnamefont {Reiter}}, \bibinfo
  {author} {\bibfnamefont {E.}~\bibnamefont {Kingma}}, \bibinfo {author}
  {\bibfnamefont {E.}~\bibnamefont {Frey}}, \ and\ \bibinfo {author}
  {\bibfnamefont {C.}~\bibnamefont {Dekker}},\ }\href {\doibase
  10.15252/msb.20156724} {\bibfield  {journal} {\bibinfo  {journal} {Molecular
  Systems Biology}\ }\textbf {\bibinfo {volume} {12}},\ \bibinfo {pages} {873}
  (\bibinfo {year} {2016})}\BibitemShut {NoStop}%
\bibitem [{\citenamefont {Goryachev}\ and\ \citenamefont
  {Leda}(2017)}]{Goryachev.Leda2017}%
  \BibitemOpen
  \bibfield  {author} {\bibinfo {author} {\bibfnamefont {A.~B.}\ \bibnamefont
  {Goryachev}}\ and\ \bibinfo {author} {\bibfnamefont {M.}~\bibnamefont
  {Leda}},\ }\href {\doibase 10.1091/mbc.e16-10-0739} {\bibfield  {journal}
  {\bibinfo  {journal} {Molecular Biology of the Cell}\ }\textbf {\bibinfo
  {volume} {28}},\ \bibinfo {pages} {370} (\bibinfo {year} {2017})}\BibitemShut
  {NoStop}%
\bibitem [{\citenamefont {Murray}\ and\ \citenamefont
  {Sourjik}(2017)}]{Murray.Sourjik2017}%
  \BibitemOpen
  \bibfield  {author} {\bibinfo {author} {\bibfnamefont {S.~M.}\ \bibnamefont
  {Murray}}\ and\ \bibinfo {author} {\bibfnamefont {V.}~\bibnamefont
  {Sourjik}},\ }\href {\doibase 10.1038/nphys4155} {\bibfield  {journal}
  {\bibinfo  {journal} {Nature Physics}\ }\textbf {\bibinfo {volume} {13}},\
  \bibinfo {pages} {1006} (\bibinfo {year} {2017})}\BibitemShut {NoStop}%
\bibitem [{\citenamefont {Denk}\ \emph {et~al.}(2018)\citenamefont {Denk},
  \citenamefont {Kretschmer}, \citenamefont {Halatek}, \citenamefont {Hartl},
  \citenamefont {Schwille},\ and\ \citenamefont {Frey}}]{Denk.etal2018}%
  \BibitemOpen
  \bibfield  {author} {\bibinfo {author} {\bibfnamefont {J.}~\bibnamefont
  {Denk}}, \bibinfo {author} {\bibfnamefont {S.}~\bibnamefont {Kretschmer}},
  \bibinfo {author} {\bibfnamefont {J.}~\bibnamefont {Halatek}}, \bibinfo
  {author} {\bibfnamefont {C.}~\bibnamefont {Hartl}}, \bibinfo {author}
  {\bibfnamefont {P.}~\bibnamefont {Schwille}}, \ and\ \bibinfo {author}
  {\bibfnamefont {E.}~\bibnamefont {Frey}},\ }\href {\doibase
  10.1073/pnas.1719801115} {\bibfield  {journal} {\bibinfo  {journal} {Proc
  Natl Acad Sci USA}\ }\textbf {\bibinfo {volume} {115}},\ \bibinfo {pages}
  {4553} (\bibinfo {year} {2018})}\BibitemShut {NoStop}%
\bibitem [{\citenamefont {Cusseddu}\ \emph {et~al.}(2018)\citenamefont
  {Cusseddu}, \citenamefont {{Edelstein-Keshet}}, \citenamefont {Mackenzie},
  \citenamefont {Portet},\ and\ \citenamefont
  {Madzvamuse}}]{Cusseddu.etal2018}%
  \BibitemOpen
  \bibfield  {author} {\bibinfo {author} {\bibfnamefont {D.}~\bibnamefont
  {Cusseddu}}, \bibinfo {author} {\bibfnamefont {L.}~\bibnamefont
  {{Edelstein-Keshet}}}, \bibinfo {author} {\bibfnamefont {J.}~\bibnamefont
  {Mackenzie}}, \bibinfo {author} {\bibfnamefont {S.}~\bibnamefont {Portet}}, \
  and\ \bibinfo {author} {\bibfnamefont {A.}~\bibnamefont {Madzvamuse}},\
  }\href {\doibase 10.1016/j.jtbi.2018.09.008} {\bibfield  {journal} {\bibinfo
  {journal} {Journal of Theoretical Biology}\ } (\bibinfo {year} {2018}),\
  10.1016/j.jtbi.2018.09.008}\BibitemShut {NoStop}%
\bibitem [{\citenamefont {Chiou}\ \emph {et~al.}(2018)\citenamefont {Chiou},
  \citenamefont {Ramirez}, \citenamefont {Elston}, \citenamefont {Witelski},
  \citenamefont {Schaeffer},\ and\ \citenamefont {Lew}}]{Chiou.etal2018}%
  \BibitemOpen
  \bibfield  {author} {\bibinfo {author} {\bibfnamefont {J.-G.}\ \bibnamefont
  {Chiou}}, \bibinfo {author} {\bibfnamefont {S.~A.}\ \bibnamefont {Ramirez}},
  \bibinfo {author} {\bibfnamefont {T.~C.}\ \bibnamefont {Elston}}, \bibinfo
  {author} {\bibfnamefont {T.~P.}\ \bibnamefont {Witelski}}, \bibinfo {author}
  {\bibfnamefont {D.~G.}\ \bibnamefont {Schaeffer}}, \ and\ \bibinfo {author}
  {\bibfnamefont {D.~J.}\ \bibnamefont {Lew}},\ }\href {\doibase
  10.1371/journal.pcbi.1006095} {\bibfield  {journal} {\bibinfo  {journal}
  {PLOS Computational Biology}\ }\textbf {\bibinfo {volume} {14}},\ \bibinfo
  {pages} {e1006095} (\bibinfo {year} {2018})}\BibitemShut {NoStop}%
\bibitem [{\citenamefont {Glock}\ \emph {et~al.}(2019)\citenamefont {Glock},
  \citenamefont {Brauns}, \citenamefont {Halatek}, \citenamefont {Frey},\ and\
  \citenamefont {Schwille}}]{Glock.etal2019}%
  \BibitemOpen
  \bibfield  {author} {\bibinfo {author} {\bibfnamefont {P.}~\bibnamefont
  {Glock}}, \bibinfo {author} {\bibfnamefont {F.}~\bibnamefont {Brauns}},
  \bibinfo {author} {\bibfnamefont {J.}~\bibnamefont {Halatek}}, \bibinfo
  {author} {\bibfnamefont {E.}~\bibnamefont {Frey}}, \ and\ \bibinfo {author}
  {\bibfnamefont {P.}~\bibnamefont {Schwille}},\ }\href {\doibase
  10.1101/666362} {\bibfield  {journal} {\bibinfo  {journal} {bioRxiv}\ }
  (\bibinfo {year} {2019}),\ 10.1101/666362}\BibitemShut {NoStop}%
\bibitem [{\citenamefont {Turing}(1952)}]{Turing1952}%
  \BibitemOpen
  \bibfield  {author} {\bibinfo {author} {\bibfnamefont {A.~M.}\ \bibnamefont
  {Turing}},\ }\href {\doibase 10.1098/rstb.1952.0012} {\bibfield  {journal}
  {\bibinfo  {journal} {Philosophical Transactions of the Royal Society of
  London. Series B, Biological Sciences}\ }\textbf {\bibinfo {volume} {237}},\
  \bibinfo {pages} {37} (\bibinfo {year} {1952})}\BibitemShut {NoStop}%
\bibitem [{\citenamefont {Williams}\ \emph {et~al.}(2019)\citenamefont
  {Williams}, \citenamefont {Paschke},\ and\ \citenamefont
  {Kay}}]{Williams.etal2019}%
  \BibitemOpen
  \bibfield  {author} {\bibinfo {author} {\bibfnamefont {T.~D.}\ \bibnamefont
  {Williams}}, \bibinfo {author} {\bibfnamefont {P.~I.}\ \bibnamefont
  {Paschke}}, \ and\ \bibinfo {author} {\bibfnamefont {R.~R.}\ \bibnamefont
  {Kay}},\ }\href {\doibase 10.1098/rstb.2018.0150} {\bibfield  {journal}
  {\bibinfo  {journal} {Philosophical Transactions of the Royal Society B:
  Biological Sciences}\ }\textbf {\bibinfo {volume} {374}},\ \bibinfo {pages}
  {20180150} (\bibinfo {year} {2019})}\BibitemShut {NoStop}%
\bibitem [{\citenamefont {Veltman}\ \emph {et~al.}(2016)\citenamefont
  {Veltman}, \citenamefont {Williams}, \citenamefont {Bloomfield},
  \citenamefont {Chen}, \citenamefont {Betzig}, \citenamefont {Insall},\ and\
  \citenamefont {Kay}}]{Veltman.etal2016}%
  \BibitemOpen
  \bibfield  {author} {\bibinfo {author} {\bibfnamefont {D.~M.}\ \bibnamefont
  {Veltman}}, \bibinfo {author} {\bibfnamefont {T.~D.}\ \bibnamefont
  {Williams}}, \bibinfo {author} {\bibfnamefont {G.}~\bibnamefont
  {Bloomfield}}, \bibinfo {author} {\bibfnamefont {B.-C.}\ \bibnamefont
  {Chen}}, \bibinfo {author} {\bibfnamefont {E.}~\bibnamefont {Betzig}},
  \bibinfo {author} {\bibfnamefont {R.~H.}\ \bibnamefont {Insall}}, \ and\
  \bibinfo {author} {\bibfnamefont {R.~R.}\ \bibnamefont {Kay}},\ }\href
  {\doibase 10.7554/eLife.20085} {\bibfield  {journal} {\bibinfo  {journal}
  {eLife}\ }\textbf {\bibinfo {volume} {5}},\ \bibinfo {pages} {e20085}
  (\bibinfo {year} {2016})}\BibitemShut {NoStop}%
\bibitem [{\citenamefont {Vaughan}\ \emph {et~al.}(2011)\citenamefont
  {Vaughan}, \citenamefont {Miller}, \citenamefont {Yu},\ and\ \citenamefont
  {Bement}}]{Vaughan.etal2011}%
  \BibitemOpen
  \bibfield  {author} {\bibinfo {author} {\bibfnamefont {E.~M.}\ \bibnamefont
  {Vaughan}}, \bibinfo {author} {\bibfnamefont {A.~L.}\ \bibnamefont {Miller}},
  \bibinfo {author} {\bibfnamefont {H.-Y.~E.}\ \bibnamefont {Yu}}, \ and\
  \bibinfo {author} {\bibfnamefont {W.~M.}\ \bibnamefont {Bement}},\ }\href
  {\doibase 10.1016/j.cub.2011.01.014} {\bibfield  {journal} {\bibinfo
  {journal} {Current Biology}\ }\textbf {\bibinfo {volume} {21}},\ \bibinfo
  {pages} {270} (\bibinfo {year} {2011})}\BibitemShut {NoStop}%
\bibitem [{\citenamefont {Chant}\ and\ \citenamefont
  {Pringle}(1995)}]{Chant.Pringle1995}%
  \BibitemOpen
  \bibfield  {author} {\bibinfo {author} {\bibfnamefont {J.}~\bibnamefont
  {Chant}}\ and\ \bibinfo {author} {\bibfnamefont {J.~R.}\ \bibnamefont
  {Pringle}},\ }\href {\doibase 10.1083/jcb.129.3.751} {\bibfield  {journal}
  {\bibinfo  {journal} {J. Cell Biol.}\ }\textbf {\bibinfo {volume} {129}},\
  \bibinfo {pages} {751} (\bibinfo {year} {1995})}\BibitemShut {NoStop}%
\bibitem [{\citenamefont {Tong}\ \emph {et~al.}(2007)\citenamefont {Tong},
  \citenamefont {Gao}, \citenamefont {Howell}, \citenamefont {Bose},
  \citenamefont {Lew},\ and\ \citenamefont {Bi}}]{Tong.etal2007}%
  \BibitemOpen
  \bibfield  {author} {\bibinfo {author} {\bibfnamefont {Z.}~\bibnamefont
  {Tong}}, \bibinfo {author} {\bibfnamefont {X.-D.}\ \bibnamefont {Gao}},
  \bibinfo {author} {\bibfnamefont {A.~S.}\ \bibnamefont {Howell}}, \bibinfo
  {author} {\bibfnamefont {I.}~\bibnamefont {Bose}}, \bibinfo {author}
  {\bibfnamefont {D.~J.}\ \bibnamefont {Lew}}, \ and\ \bibinfo {author}
  {\bibfnamefont {E.}~\bibnamefont {Bi}},\ }\href {\doibase
  10.1083/jcb.200705160} {\bibfield  {journal} {\bibinfo  {journal} {J Cell
  Biol}\ }\textbf {\bibinfo {volume} {179}},\ \bibinfo {pages} {1375} (\bibinfo
  {year} {2007})}\BibitemShut {NoStop}%
\bibitem [{\citenamefont {Lo}\ \emph {et~al.}(2013)\citenamefont {Lo},
  \citenamefont {Lee}, \citenamefont {Narayan}, \citenamefont {Chou},\ and\
  \citenamefont {Park}}]{Lo.etal2013}%
  \BibitemOpen
  \bibfield  {author} {\bibinfo {author} {\bibfnamefont {W.-C.}\ \bibnamefont
  {Lo}}, \bibinfo {author} {\bibfnamefont {M.~E.}\ \bibnamefont {Lee}},
  \bibinfo {author} {\bibfnamefont {M.}~\bibnamefont {Narayan}}, \bibinfo
  {author} {\bibfnamefont {C.-S.}\ \bibnamefont {Chou}}, \ and\ \bibinfo
  {author} {\bibfnamefont {H.-O.}\ \bibnamefont {Park}},\ }\href {\doibase
  10.1371/journal.pone.0056665} {\bibfield  {journal} {\bibinfo  {journal}
  {PLoS ONE}\ }\textbf {\bibinfo {volume} {8}},\ \bibinfo {pages} {e56665}
  (\bibinfo {year} {2013})}\BibitemShut {NoStop}%
\bibitem [{\citenamefont {Miller}\ \emph {et~al.}(2017)\citenamefont {Miller},
  \citenamefont {Lo}, \citenamefont {Lee}, \citenamefont {Kang},\ and\
  \citenamefont {Park}}]{Miller.etal2017}%
  \BibitemOpen
  \bibfield  {author} {\bibinfo {author} {\bibfnamefont {K.~E.}\ \bibnamefont
  {Miller}}, \bibinfo {author} {\bibfnamefont {W.-C.}\ \bibnamefont {Lo}},
  \bibinfo {author} {\bibfnamefont {M.~E.}\ \bibnamefont {Lee}}, \bibinfo
  {author} {\bibfnamefont {P.~J.}\ \bibnamefont {Kang}}, \ and\ \bibinfo
  {author} {\bibfnamefont {H.-O.}\ \bibnamefont {Park}},\ }\href {\doibase
  10.1091/mbc.e17-01-0074} {\bibfield  {journal} {\bibinfo  {journal}
  {Molecular Biology of the Cell}\ }\textbf {\bibinfo {volume} {28}},\ \bibinfo
  {pages} {3773} (\bibinfo {year} {2017})}\BibitemShut {NoStop}%
\bibitem [{\citenamefont {Halatek}\ and\ \citenamefont
  {Frey}(2018)}]{Halatek.Frey2018}%
  \BibitemOpen
  \bibfield  {author} {\bibinfo {author} {\bibfnamefont {J.}~\bibnamefont
  {Halatek}}\ and\ \bibinfo {author} {\bibfnamefont {E.}~\bibnamefont {Frey}},\
  }\href {\doibase 10.1038/s41567-017-0040-5} {\bibfield  {journal} {\bibinfo
  {journal} {Nature Physics}\ }\textbf {\bibinfo {volume} {14}},\ \bibinfo
  {pages} {507} (\bibinfo {year} {2018})}\BibitemShut {NoStop}%
\bibitem [{\citenamefont {Brauns}\ \emph {et~al.}(2018)\citenamefont {Brauns},
  \citenamefont {Halatek},\ and\ \citenamefont {Frey}}]{Brauns.etal2018}%
  \BibitemOpen
  \bibfield  {author} {\bibinfo {author} {\bibfnamefont {F.}~\bibnamefont
  {Brauns}}, \bibinfo {author} {\bibfnamefont {J.}~\bibnamefont {Halatek}}, \
  and\ \bibinfo {author} {\bibfnamefont {E.}~\bibnamefont {Frey}},\ }\href@noop
  {} {\bibfield  {journal} {\bibinfo  {journal} {arXiv:1812.08684}\ } (\bibinfo
  {year} {2018})},\ \Eprint {http://arxiv.org/abs/1812.08684}
  {arXiv:1812.08684} \BibitemShut {NoStop}%
\bibitem [{\citenamefont {Maini}\ \emph {et~al.}(1992)\citenamefont {Maini},
  \citenamefont {Benson},\ and\ \citenamefont {Sherratt}}]{Maini.etal1992}%
  \BibitemOpen
  \bibfield  {author} {\bibinfo {author} {\bibfnamefont {P.~K.}\ \bibnamefont
  {Maini}}, \bibinfo {author} {\bibfnamefont {D.~L.}\ \bibnamefont {Benson}}, \
  and\ \bibinfo {author} {\bibfnamefont {J.~A.}\ \bibnamefont {Sherratt}},\
  }\href {\doibase 10.1093/imammb/9.3.197} {\bibfield  {journal} {\bibinfo
  {journal} {Math Med Biol}\ }\textbf {\bibinfo {volume} {9}},\ \bibinfo
  {pages} {197} (\bibinfo {year} {1992})}\BibitemShut {NoStop}%
\bibitem [{\citenamefont {Ermentrout}\ and\ \citenamefont
  {Rinzel}(1996)}]{Ermentrout.Rinzel1996}%
  \BibitemOpen
  \bibfield  {author} {\bibinfo {author} {\bibfnamefont {G.~B.}\ \bibnamefont
  {Ermentrout}}\ and\ \bibinfo {author} {\bibfnamefont {J.}~\bibnamefont
  {Rinzel}},\ }\href {\doibase 10.1137/S0036139994276793} {\bibfield  {journal}
  {\bibinfo  {journal} {SIAM Journal on Applied Mathematics}\ }\textbf
  {\bibinfo {volume} {56}},\ \bibinfo {pages} {1107} (\bibinfo {year}
  {1996})}\BibitemShut {NoStop}%
\bibitem [{\citenamefont {B{\"a}r}\ \emph {et~al.}(1996)\citenamefont
  {B{\"a}r}, \citenamefont {Bangia}, \citenamefont {Kevrekidis}, \citenamefont
  {Haas}, \citenamefont {Rotermund},\ and\ \citenamefont
  {Ertl}}]{Bar.etal1996}%
  \BibitemOpen
  \bibfield  {author} {\bibinfo {author} {\bibfnamefont {M.}~\bibnamefont
  {B{\"a}r}}, \bibinfo {author} {\bibfnamefont {A.~K.}\ \bibnamefont {Bangia}},
  \bibinfo {author} {\bibfnamefont {I.~G.}\ \bibnamefont {Kevrekidis}},
  \bibinfo {author} {\bibfnamefont {G.}~\bibnamefont {Haas}}, \bibinfo {author}
  {\bibfnamefont {H.-H.}\ \bibnamefont {Rotermund}}, \ and\ \bibinfo {author}
  {\bibfnamefont {G.}~\bibnamefont {Ertl}},\ }\href {\doibase
  10.1021/jp961689q} {\bibfield  {journal} {\bibinfo  {journal} {The Journal of
  Physical Chemistry}\ }\textbf {\bibinfo {volume} {100}},\ \bibinfo {pages}
  {19106} (\bibinfo {year} {1996})}\BibitemShut {NoStop}%
\bibitem [{\citenamefont {Yuan}\ \emph {et~al.}(2007)\citenamefont {Yuan},
  \citenamefont {Teramoto},\ and\ \citenamefont {Nishiura}}]{Yuan.etal2007}%
  \BibitemOpen
  \bibfield  {author} {\bibinfo {author} {\bibfnamefont {X.}~\bibnamefont
  {Yuan}}, \bibinfo {author} {\bibfnamefont {T.}~\bibnamefont {Teramoto}}, \
  and\ \bibinfo {author} {\bibfnamefont {Y.}~\bibnamefont {Nishiura}},\ }\href
  {\doibase 10.1103/PhysRevE.75.036220} {\bibfield  {journal} {\bibinfo
  {journal} {Phys. Rev. E}\ }\textbf {\bibinfo {volume} {75}},\ \bibinfo
  {pages} {036220} (\bibinfo {year} {2007})}\BibitemShut {NoStop}%
\bibitem [{\citenamefont {Nishiura}\ \emph {et~al.}(2007)\citenamefont
  {Nishiura}, \citenamefont {Teramoto}, \citenamefont {Yuan},\ and\
  \citenamefont {Ueda}}]{Nishiura.etal2007}%
  \BibitemOpen
  \bibfield  {author} {\bibinfo {author} {\bibfnamefont {Y.}~\bibnamefont
  {Nishiura}}, \bibinfo {author} {\bibfnamefont {T.}~\bibnamefont {Teramoto}},
  \bibinfo {author} {\bibfnamefont {X.}~\bibnamefont {Yuan}}, \ and\ \bibinfo
  {author} {\bibfnamefont {K.-I.}\ \bibnamefont {Ueda}},\ }\href {\doibase
  10.1063/1.2778553} {\bibfield  {journal} {\bibinfo  {journal} {Chaos: An
  Interdisciplinary Journal of Nonlinear Science}\ }\textbf {\bibinfo {volume}
  {17}},\ \bibinfo {pages} {037104} (\bibinfo {year} {2007})}\BibitemShut
  {NoStop}%
\bibitem [{\citenamefont {Miyazaki}\ and\ \citenamefont
  {Kinoshita}(2007)}]{Miyazaki.Kinoshita2007}%
  \BibitemOpen
  \bibfield  {author} {\bibinfo {author} {\bibfnamefont {J.}~\bibnamefont
  {Miyazaki}}\ and\ \bibinfo {author} {\bibfnamefont {S.}~\bibnamefont
  {Kinoshita}},\ }\href {\doibase 10.1103/PhysRevE.76.066201} {\bibfield
  {journal} {\bibinfo  {journal} {Physical Review E}\ }\textbf {\bibinfo
  {volume} {76}} (\bibinfo {year} {2007}),\
  10.1103/PhysRevE.76.066201}\BibitemShut {NoStop}%
\bibitem [{\citenamefont {Nishi}\ \emph {et~al.}(2013)\citenamefont {Nishi},
  \citenamefont {Nishiura},\ and\ \citenamefont {Teramoto}}]{Nishi.etal2013}%
  \BibitemOpen
  \bibfield  {author} {\bibinfo {author} {\bibfnamefont {K.}~\bibnamefont
  {Nishi}}, \bibinfo {author} {\bibfnamefont {Y.}~\bibnamefont {Nishiura}}, \
  and\ \bibinfo {author} {\bibfnamefont {T.}~\bibnamefont {Teramoto}},\ }\href
  {\doibase 10.1007/s13160-013-0100-x} {\bibfield  {journal} {\bibinfo
  {journal} {Japan Journal of Industrial and Applied Mathematics}\ }\textbf
  {\bibinfo {volume} {30}},\ \bibinfo {pages} {351} (\bibinfo {year}
  {2013})}\BibitemShut {NoStop}%
\bibitem [{\citenamefont {Doelman}\ \emph {et~al.}(2016)\citenamefont
  {Doelman}, \citenamefont {{van Heijster}},\ and\ \citenamefont
  {Xie}}]{Doelman.etal2016}%
  \BibitemOpen
  \bibfield  {author} {\bibinfo {author} {\bibfnamefont {A.}~\bibnamefont
  {Doelman}}, \bibinfo {author} {\bibfnamefont {P.}~\bibnamefont {{van
  Heijster}}}, \ and\ \bibinfo {author} {\bibfnamefont {F.}~\bibnamefont
  {Xie}},\ }\href {\doibase 10.1137/15M1026742} {\bibfield  {journal} {\bibinfo
   {journal} {SIAM Journal on Applied Dynamical Systems}\ }\textbf {\bibinfo
  {volume} {15}},\ \bibinfo {pages} {655} (\bibinfo {year} {2016})}\BibitemShut
  {NoStop}%
\bibitem [{\citenamefont {{van Heijster}}\ \emph {et~al.}(2019)\citenamefont
  {{van Heijster}}, \citenamefont {Chen}, \citenamefont {Nishiura},\ and\
  \citenamefont {Teramoto}}]{vanHeijster.etal2019}%
  \BibitemOpen
  \bibfield  {author} {\bibinfo {author} {\bibfnamefont {P.}~\bibnamefont {{van
  Heijster}}}, \bibinfo {author} {\bibfnamefont {C.-N.}\ \bibnamefont {Chen}},
  \bibinfo {author} {\bibfnamefont {Y.}~\bibnamefont {Nishiura}}, \ and\
  \bibinfo {author} {\bibfnamefont {T.}~\bibnamefont {Teramoto}},\ }\href
  {\doibase 10.1007/s10884-018-9694-7} {\bibfield  {journal} {\bibinfo
  {journal} {J Dyn Diff Equat}\ }\textbf {\bibinfo {volume} {31}},\ \bibinfo
  {pages} {153} (\bibinfo {year} {2019})}\BibitemShut {NoStop}%
\bibitem [{\citenamefont {Scheel}\ and\ \citenamefont
  {Weinburd}(2018)}]{Scheel.Weinburd2018}%
  \BibitemOpen
  \bibfield  {author} {\bibinfo {author} {\bibfnamefont {A.}~\bibnamefont
  {Scheel}}\ and\ \bibinfo {author} {\bibfnamefont {J.}~\bibnamefont
  {Weinburd}},\ }\href {\doibase 10.1098/rsta.2017.0191} {\bibfield  {journal}
  {\bibinfo  {journal} {Philosophical Transactions of the Royal Society A:
  Mathematical, Physical and Engineering Sciences}\ }\textbf {\bibinfo {volume}
  {376}},\ \bibinfo {pages} {20170191} (\bibinfo {year} {2018})}\BibitemShut
  {NoStop}%
\bibitem [{\citenamefont {Jones}(1995)}]{Jones1995}%
  \BibitemOpen
  \bibfield  {author} {\bibinfo {author} {\bibfnamefont {C.~K. R.~T.}\
  \bibnamefont {Jones}},\ }in\ \href {\doibase 10.1007/BFb0095239} {\emph
  {\bibinfo {booktitle} {Dynamical {{Systems}}}}},\ Vol.\ \bibinfo {volume}
  {1609},\ \bibinfo {editor} {edited by\ \bibinfo {editor} {\bibfnamefont
  {R.}~\bibnamefont {Johnson}}}\ (\bibinfo  {publisher} {{Springer}},\ \bibinfo
  {address} {{Berlin, Heidelberg}},\ \bibinfo {year} {1995})\ pp.\ \bibinfo
  {pages} {44--118}\BibitemShut {NoStop}%
\bibitem [{\citenamefont {Ward}(2006)}]{Ward2006}%
  \BibitemOpen
  \bibfield  {author} {\bibinfo {author} {\bibfnamefont {M.~J.}\ \bibnamefont
  {Ward}},\ }\href {\doibase 10.1007/s11538-006-9091-y} {\bibfield  {journal}
  {\bibinfo  {journal} {Bulletin of Mathematical Biology}\ }\textbf {\bibinfo
  {volume} {68}},\ \bibinfo {pages} {1151} (\bibinfo {year}
  {2006})}\BibitemShut {NoStop}%
\bibitem [{\citenamefont {Guckenheimer}\ and\ \citenamefont
  {Holmes}(1983)}]{Guckenheimer.Holmes1983}%
  \BibitemOpen
  \bibfield  {author} {\bibinfo {author} {\bibfnamefont {J.}~\bibnamefont
  {Guckenheimer}}\ and\ \bibinfo {author} {\bibfnamefont {P.}~\bibnamefont
  {Holmes}},\ }\href {\doibase 10.1007/978-1-4612-1140-2} {\emph {\bibinfo
  {title} {Nonlinear {{Oscillations}}, {{Dynamical Systems}}, and
  {{Bifurcations}} of {{Vector Fields}}}}},\ \bibinfo {series} {Applied
  {{Mathematical Sciences}}}, Vol.~\bibinfo {volume} {42}\ (\bibinfo
  {publisher} {{Springer New York}},\ \bibinfo {address} {{New York, NY}},\
  \bibinfo {year} {1983})\BibitemShut {NoStop}%
\bibitem [{\citenamefont {Krauskopf}\ \emph {et~al.}(2007)\citenamefont
  {Krauskopf}, \citenamefont {Osinga},\ and\ \citenamefont
  {{Galan-Vioque}}}]{Krauskopf.etal2007}%
  \BibitemOpen
  \bibinfo {editor} {\bibfnamefont {B.}~\bibnamefont {Krauskopf}}, \bibinfo
  {editor} {\bibfnamefont {H.~M.}\ \bibnamefont {Osinga}}, \ and\ \bibinfo
  {editor} {\bibfnamefont {J.}~\bibnamefont {{Galan-Vioque}}},\ eds.,\ \href
  {\doibase 10.1007/978-1-4020-6356-5} {\emph {\bibinfo {title} {Numerical
  {{Continuation Methods}} for {{Dynamical Systems}}: {{Path}} Following and
  Boundary Value Problems}}},\ Understanding {{Complex Systems}}\ (\bibinfo
  {publisher} {{Springer Netherlands}},\ \bibinfo {address} {{Dordrecht}},\
  \bibinfo {year} {2007})\BibitemShut {NoStop}%
\bibitem [{\citenamefont {Pryce}(1993)}]{Pryce1993}%
  \BibitemOpen
  \bibfield  {author} {\bibinfo {author} {\bibfnamefont {J.~D.}\ \bibnamefont
  {Pryce}},\ }\href@noop {} {\emph {\bibinfo {title} {Numerical Solution of
  {{Sturm}}-{{Liouville}} Problems}}},\ Monographs on Numerical Analysis\
  (\bibinfo  {publisher} {{Clarendon Press}},\ \bibinfo {address} {{Oxford; New
  York}},\ \bibinfo {year} {1993})\BibitemShut {NoStop}%
\bibitem [{\citenamefont {Ziepke}\ \emph {et~al.}(2016)\citenamefont {Ziepke},
  \citenamefont {Martens},\ and\ \citenamefont {Engel}}]{Ziepke.etal2016}%
  \BibitemOpen
  \bibfield  {author} {\bibinfo {author} {\bibfnamefont {A.}~\bibnamefont
  {Ziepke}}, \bibinfo {author} {\bibfnamefont {S.}~\bibnamefont {Martens}}, \
  and\ \bibinfo {author} {\bibfnamefont {H.}~\bibnamefont {Engel}},\ }\href
  {\doibase 10.1063/1.4962173} {\bibfield  {journal} {\bibinfo  {journal} {J.
  Chem. Phys.}\ }\textbf {\bibinfo {volume} {145}},\ \bibinfo {pages} {094108}
  (\bibinfo {year} {2016})}\BibitemShut {NoStop}%
\bibitem [{\citenamefont {Pfeffer}(2012)}]{Pfeffer2012}%
  \BibitemOpen
  \bibfield  {author} {\bibinfo {author} {\bibfnamefont {S.~R.}\ \bibnamefont
  {Pfeffer}},\ }\href {\doibase 10.1042/BST20120168} {\bibfield  {journal}
  {\bibinfo  {journal} {Biochemical Society Transactions}\ }\textbf {\bibinfo
  {volume} {40}},\ \bibinfo {pages} {1373} (\bibinfo {year}
  {2012})}\BibitemShut {NoStop}%
\bibitem [{\citenamefont {Novick}(2016)}]{Novick2016}%
  \BibitemOpen
  \bibfield  {author} {\bibinfo {author} {\bibfnamefont {P.}~\bibnamefont
  {Novick}},\ }\href {\doibase 10.1080/21541248.2016.1213781} {\bibfield
  {journal} {\bibinfo  {journal} {Small GTPases}\ }\textbf {\bibinfo {volume}
  {7}},\ \bibinfo {pages} {252} (\bibinfo {year} {2016})}\BibitemShut {NoStop}%
\bibitem [{\citenamefont {Noack}\ and\ \citenamefont
  {Jaillais}(2017)}]{Noack.Jaillais2017}%
  \BibitemOpen
  \bibfield  {author} {\bibinfo {author} {\bibfnamefont {L.~C.}\ \bibnamefont
  {Noack}}\ and\ \bibinfo {author} {\bibfnamefont {Y.}~\bibnamefont
  {Jaillais}},\ }\href {\doibase 10.1016/j.pbi.2017.06.017} {\bibfield
  {journal} {\bibinfo  {journal} {Current Opinion in Plant Biology}\ }\textbf
  {\bibinfo {volume} {40}},\ \bibinfo {pages} {22} (\bibinfo {year}
  {2017})}\BibitemShut {NoStop}%
\bibitem [{\citenamefont {Rulands}\ \emph {et~al.}(2013)\citenamefont
  {Rulands}, \citenamefont {Kl{\"u}nder},\ and\ \citenamefont
  {Frey}}]{Rulands.etal2013}%
  \BibitemOpen
  \bibfield  {author} {\bibinfo {author} {\bibfnamefont {S.}~\bibnamefont
  {Rulands}}, \bibinfo {author} {\bibfnamefont {B.}~\bibnamefont
  {Kl{\"u}nder}}, \ and\ \bibinfo {author} {\bibfnamefont {E.}~\bibnamefont
  {Frey}},\ }\href {\doibase 10.1103/PhysRevLett.110.038102} {\bibfield
  {journal} {\bibinfo  {journal} {Phys. Rev. Lett.}\ }\textbf {\bibinfo
  {volume} {110}},\ \bibinfo {pages} {038102} (\bibinfo {year}
  {2013})}\BibitemShut {NoStop}%
\bibitem [{\citenamefont {Zadorin}\ \emph {et~al.}(2017)\citenamefont
  {Zadorin}, \citenamefont {Rondelez}, \citenamefont {Gines}, \citenamefont
  {Dilhas}, \citenamefont {Urtel}, \citenamefont {Zambrano}, \citenamefont
  {Galas},\ and\ \citenamefont {{Estevez-Torres}}}]{Zadorin.etal2017}%
  \BibitemOpen
  \bibfield  {author} {\bibinfo {author} {\bibfnamefont {A.~S.}\ \bibnamefont
  {Zadorin}}, \bibinfo {author} {\bibfnamefont {Y.}~\bibnamefont {Rondelez}},
  \bibinfo {author} {\bibfnamefont {G.}~\bibnamefont {Gines}}, \bibinfo
  {author} {\bibfnamefont {V.}~\bibnamefont {Dilhas}}, \bibinfo {author}
  {\bibfnamefont {G.}~\bibnamefont {Urtel}}, \bibinfo {author} {\bibfnamefont
  {A.}~\bibnamefont {Zambrano}}, \bibinfo {author} {\bibfnamefont {J.-C.}\
  \bibnamefont {Galas}}, \ and\ \bibinfo {author} {\bibfnamefont
  {A.}~\bibnamefont {{Estevez-Torres}}},\ }\href {\doibase 10.1038/nchem.2770}
  {\bibfield  {journal} {\bibinfo  {journal} {Nature Chemistry}\ }\textbf
  {\bibinfo {volume} {9}},\ \bibinfo {pages} {990} (\bibinfo {year}
  {2017})}\BibitemShut {NoStop}%
\bibitem [{\citenamefont {Thalmeier}\ \emph {et~al.}(2016)\citenamefont
  {Thalmeier}, \citenamefont {Halatek},\ and\ \citenamefont
  {Frey}}]{Thalmeier.etal2016}%
  \BibitemOpen
  \bibfield  {author} {\bibinfo {author} {\bibfnamefont {D.}~\bibnamefont
  {Thalmeier}}, \bibinfo {author} {\bibfnamefont {J.}~\bibnamefont {Halatek}},
  \ and\ \bibinfo {author} {\bibfnamefont {E.}~\bibnamefont {Frey}},\ }\href
  {\doibase 10.1073/pnas.1515191113} {\bibfield  {journal} {\bibinfo  {journal}
  {Proc Natl Acad Sci USA}\ }\textbf {\bibinfo {volume} {113}},\ \bibinfo
  {pages} {548} (\bibinfo {year} {2016})}\BibitemShut {NoStop}%
\bibitem [{\citenamefont {Ziepke}\ \emph {et~al.}(2019)\citenamefont {Ziepke},
  \citenamefont {Martens},\ and\ \citenamefont {Engel}}]{Ziepke.etal2019}%
  \BibitemOpen
  \bibfield  {author} {\bibinfo {author} {\bibfnamefont {A.}~\bibnamefont
  {Ziepke}}, \bibinfo {author} {\bibfnamefont {S.}~\bibnamefont {Martens}}, \
  and\ \bibinfo {author} {\bibfnamefont {H.}~\bibnamefont {Engel}},\ }\href
  {\doibase 10.1137/18M1197278} {\bibfield  {journal} {\bibinfo  {journal}
  {SIAM J. Appl. Dyn. Syst.}\ }\textbf {\bibinfo {volume} {18}},\ \bibinfo
  {pages} {1015} (\bibinfo {year} {2019})}\BibitemShut {NoStop}%
\bibitem [{\citenamefont {Haupt}\ and\ \citenamefont
  {Minc}(2018)}]{Haupt.Minc2018}%
  \BibitemOpen
  \bibfield  {author} {\bibinfo {author} {\bibfnamefont {A.}~\bibnamefont
  {Haupt}}\ and\ \bibinfo {author} {\bibfnamefont {N.}~\bibnamefont {Minc}},\
  }\href {\doibase 10.1242/jcs.214015} {\bibfield  {journal} {\bibinfo
  {journal} {Journal of Cell Science}\ }\textbf {\bibinfo {volume} {131}},\
  \bibinfo {pages} {jcs214015} (\bibinfo {year} {2018})}\BibitemShut {NoStop}%
\bibitem [{\citenamefont {{van Saarloos}}(2003)}]{vanSaarloos2003}%
  \BibitemOpen
  \bibfield  {author} {\bibinfo {author} {\bibfnamefont {W.}~\bibnamefont {{van
  Saarloos}}},\ }\href {\doibase 10.1016/j.physrep.2003.08.001} {\bibfield
  {journal} {\bibinfo  {journal} {Physics Reports}\ }\textbf {\bibinfo {volume}
  {386}},\ \bibinfo {pages} {29} (\bibinfo {year} {2003})}\BibitemShut
  {NoStop}%
\end{thebibliography}
%

\end{document}